\documentclass[showkeys,prd,nofootinbib,floatfix,preprintnumbers]{revtex4-1}

\usepackage[utf8]{inputenc}
\usepackage{multirow}
\usepackage{amsfonts,amsmath,amssymb} 
\usepackage{graphicx,graphics,color}
\usepackage{gensymb}
\usepackage{longtable}
\usepackage{bbding}
\usepackage{subfigure}
\usepackage[T1]{fontenc}
\usepackage{hyperref}
\usepackage[normalem]{ulem}
\hypersetup{
    colorlinks=true,
    linkcolor=blue}

\begin{document}




\title{Viscous Unified Dark Matter Models Under Scrutiny: \\ Uncovering Inconsistencies from Dynamical System Analysis}

\author{Guillermo Palma}
\email{guillermo.palma@usach.cl}
\affiliation{Departamento de F\'isica, Universidad de Santiago de Chile,\\Avenida V\'ictor Jara 3493, Estaci\'on Central, 9170124, Santiago, Chile}

\author{Gabriel G\'omez}
\email{gabriel.gomez.d@usach.cl}
\affiliation{Departamento de F\'isica, Universidad de Santiago de Chile,\\Avenida V\'ictor Jara 3493, Estaci\'on Central, 9170124, Santiago, Chile}

\author{Norman Cruz}
\email{norman.cruz@usach.cl}
\affiliation{Departamento de F\'isica, Universidad de Santiago de Chile,\\Avenida V\'ictor Jara 3493, Estaci\'on Central, 9170124, Santiago, Chile}


\date{\today}

\begin{abstract}
Viscous unified dark matter models aim to describe the dark sector of the Universe, 
as the dark matter fluid itself gives rise to an accelerated expansion due to a negative bulk viscous pressure. However, in most studies, radiation is often disregarded as a minor factor in dynamical system analyses, overlooking whether radiation domination is achievable. In this paper, we rigorously examine this critical aspect for common parameterizations of bulk viscosity, denoted as $\xi$, within two general classes of viscous unified models. Our findings reveal significant inconsistencies in models where $\xi \propto H^{1-2s} \rho_{m}^{s}$ with $s\leq 0$, and surprisingly, in models where $\xi \propto \rho_{m}^{s}$ with exponents $s<0$, as they both fail to produce a radiation-dominated era—an undesirable outcome for any cosmological model. Moreover, the exponent $s$ must lye within the interval $0 \leq s < 1/2$ for the latter model to correctly describes the cosmological evolution. 
These results underscore the need of including these constraints as a prior in statistical analyses of observational data, with implications for current statistical inferences for the second model, where both prior and best-fit values of $s$ often fall outside the acceptable range.

\end{abstract}

\pacs{Valid PACS appear here}
\maketitle

\section{Introduction}

The study of the universe's composition and evolution has long been a central pursuit of cosmology, driving our understanding of fundamental physics \cite{Peebles:1994xt}. Among the various challenges cosmologists face, reconciling observations with theoretical models remains a constant endeavor. While the existence of dark energy is well-established from observations, its nature remains elusive \cite{Peebles:2002gy}. One promising avenue in this pursuit is the exploration of unified dark scenarios, where dark matter and dark energy are manifestations of a single phenomenon.
A promising class of unified models are viscous unified models \cite{Padmanabhan:1987dg,Bulk1,Bulk2,Colistete:2007xi,Brevik:2017msy}. These models offer an intriguing alternative, suggesting that the accelerated expansion could be a consequence of dissipative processes within the cosmic fluid, rather than the presence of the cosmological constant within the $\Lambda$CDM model. Viscous unified scenarios seek to explain a multitude of observational phenomena, ranging from the  current accelerated expansion observed in distant supernovae to large-scale structure of the universe \cite{blas2015large,Barbosa:2017ojt,foot2016solving}.


On the other hand, there are several microscopic models to explain how bulk viscosity could arise in cosmological scenarios. Among them, we mention the inclusion of self-interacting scalar fields to describe dark energy \cite{Gagnon:2011id}, within thermal field theory. Also from a microscopic point of view, the relation between particle creation and bulk viscosity in the early universe is discussed in \cite{Brevik:1996ca,Murphy:1973zz,Hu:1982uea,Eshaghi:2015tqa}, which plays an important role in the inflationary viscous model \cite{Bamba:2015sxa}. In addition to the discussion made by \cite{Gagnon:2011id}, a different microscopic model that considers a bulk viscosity induced by DM annihilation is discussed in \cite{Wilson:2006gf, Mathews:2008hk}, suited for the late time accelerated expansion.
The kinetic theory formalism has been also implemented to describe the viscous effect within self-interacting DM models \cite{Atreya:2017pny, Natwariya:2019fif}.
In the context of neutralino CDM, an energy dissipation from the CDM fluid to the radiation fluid is manifested in a collisional damping mechanism during the kinetic decoupling \cite{Hofmann:2001bi}. 

The examples mentioned above underscore the significance of taking into account different dissipative processes and their potential impacts on the behavior of the cosmic fluid. Nevertheless, there is currently no widely accepted model that incorporates a microscopically motivated bulk viscosity. This is why many viscous unified models rely on a phenomenological parametrization to describe bulk viscosity. Moreover, most of these models are framed within the Eckart's theory \cite{Eckart} (also see \cite{LandauandLifshitz}), which is a first approximation for studying relativistic and causal non-perfect fluids. We do not criticize this approach; instead, we examine in this paper their predictions and implications for cosmic dynamics using dynamical system techniques and explore potential avenues for future research. Through this exploration, we aim to provide a comprehensive understanding of viscous unified models and to uncover some inconsistencies that have been overlooked thus far. In particular, we consider two different viscous models within a unified dark matter scenario that encompass a wide range of phenomenological parametrizations for bulk viscosity. The first one is a recently proposed parametrization, which includes simultaneously the effects of the Hubble parameter and the dark matter energy density \cite{Gomez:2022qcu}\footnote{Although this model was initially proposed to describe a dissipative dark matter model as an extension of the $\Lambda$CDM, it can be naturally employed as a unified viscous model.}
\begin{equation}
    \xi^I \propto H^{1-2s} \rho_{m}^{s}. \label{sec2:eqn3}
\end{equation}
where $s$ is the viscous exponent. The above functional form has the advantage of turning on the effect of bulk viscosity when dark matter is dominant. This important feature is also satisfied by the second viscous model, which has a power law behavior in the matter energy density:
\begin{equation}
    \xi^{II} \propto \rho_{m}^{s}. \label{sec2:eqn4}
\end{equation}
The particular case $s=1/2$  has been widely investigated in bulk viscosity cosmological scenarios \cite{Big.Bang,rho1,Brevik,Analysing,valorxi1,valorxi2,primerarticulo,sym14091866,Cruz:2022zxe}. Considering both parametrizations, they offer the compelling advantage of describing well-motivated phenomenological viscous models found in the literature. 

Demands for complete cosmological dynamics, encompassing a universe transitioning from a radiation era to a decelerated matter-dominated phase, followed by an accelerated expansion, select suitable cosmological scenarios. In our context, this criterion provides valuable insights into which specific parameterizations for bulk viscosity are plausible. Once these models pass this challenge, they must undergo careful scrutiny against observational constraints, not vice versa, as is usually done. Thus, dynamical system analyses serve as an initial inquiry, indicating the values of the model parameters within the proposed parametrization that lead to successful background cosmic dynamics. It is important to note that in most studies, radiation is often disregarded as a minor factor in dynamical system analyses, overlooking whether radiation domination is achievable \cite{Coley:1995dpj,coley1996qualitative,Bulk1,Bulk4,Colistete:2007xi,acquaviva2015nonlinear,Acquaviva:2014vga}, with few exceptions \cite{Padmanabhan:1987dg,Avelino:2013wea,Avelino:2013mua,Sasidharan:2015ihq} employing different setups from ours. The exclusion of radiation is usually justified when focusing solely on late-time observations. However, we will demonstrate that the realization of a radiation-dominated phase is not a trivial aspect in viscous unified models, despite its subdominant effect at present times.

The present paper is organized as follows: After a concise introduction, we summarize the essential features of viscous unified models in Section II. Then, in Section III, we review the conditions for accelerated expansion independently of the form of the bulk viscosity coefficient. Subsequently, in Sections IV and V, we respectively perform dynamical system analyses for models I and II to establish the stability conditions of the fixed points. This is accompanied by a comprehensive analytical treatment for specific values of the exponent $s$. Finally, the main findings and a general discussion of this work are presented in Section VI.

\section{Bulk viscous Models}
In this section, we write the fundamental equations describing the dynamics of bulk viscous unified models. Within this framework, the bulk viscosity in the dark matter fluid provides a repulsive pressure capable of driving the current cosmic accelerated expansion, similar to the role played by the cosmological constant in the standard $\Lambda$CDM model. We include radiation in our prescription to illustrate later that demanding a complete cosmological evolution is not always plausible in these scenarios. We assume a homogeneous and isotropic background described by a flat FLRW metric. The matter content is represented by two fluids: one for radiation with density $\rho_{r}$ and pressure $P_{r}$, and another for dark matter with density $\rho_{m}$ and effective pressure $P_{m}^{\rm eff}$. The Friedmann equation and the acceleration equation are respectively given by: 
%
\begin{eqnarray}
    3H^{2}&=&8\pi G_{N} \left(\rho_{r}+\rho_{m}\right),  \label{sec2:eqn0}\\ 
    3H^{2}+2\dot{H}&=&-8\pi G_{N} \left(P_{r}+P_{m}^{\rm eff}\right).
    \label{sec2:eqn1}
\end{eqnarray}
Here $G_N$ denotes the Newton's constant. We use a polytropic relation for radiation $P_{r}=\rho_{r}/3$ and consider intrinsic bulk viscosity in the dark matter fluid, according to the Eckart theory, with vanishing kinetic pressure. Hence, the effective pressure is defined as
%
\begin{equation}
    P_{m}^{\rm eff}=P_{m}+\Pi=-3H\xi,\label{sec2:eqn2}
\end{equation}
and the bulk viscosity coefficient $\xi$ is positive in order to fulfill the second law of thermodynamics. Our aim is to encompass well-motivated unified bulk viscosity models within a general framework. This is achieved by employing two parametrizations for the bulk viscosity. The first parametrization corresponds to a new model recently proposed in \cite{Gomez:2022qcu} (see also \cite{Cruz:2023dzn}):
%
\begin{equation}
    \xi^I \equiv \frac{\xi_{0}}{8\pi G_{N}} H^{1-2s} H_{0}^{2s}\left(\frac{\rho_{m}}{\rho_{m,0}} \right)^{s}=\frac{\hat{\xi}_{0}}{8\pi G_{N}} H\; \Omega_{m}^{s},\label{sec2:eqn3}
\end{equation}
which possesses two significant features: firstly, the bulk viscosity effects are turned off when the energy density of matter vanishes, and secondly, it allows the dynamical system of equations (Eqns.~(\ref{sec2:eqn1})-(\ref{sec2:eqn3})) to be written as an autonomous system for arbitrary $s$-exponents, as we will demonstrate below. This parametrization is quite general, covering functional forms such as $\xi \propto H$, $\xi \propto \rho_{m}^{1/2}$, and more general combinations of both dependencies. A second viscous model is provided by the power law parametrization
\begin{equation}
    \xi^{II} \equiv \frac{\hat{\xi_{0}}/\sqrt{3}}{(8\pi G_{N})^{(3-4s)/2}} \rho_{m}^{s}. 
\label{sec2:viscous_II}
\end{equation}
This formulation includes the widely used case of a constant bulk viscosity coefficient. 
It is worthwhile to point out that both $\xi_{0}$ and $\hat{\xi}_{0}$ are dimensionless parameters independent of the $s$ value within this setup. It is also noteworthy that both parametrizations coincide for the particular value $s=1/2$. We refer to them as models I (\ref{sec2:eqn3}) and II (\ref{sec2:viscous_II}), respectively, throughout. 
To complete the dynamical system, conservation laws for the two components are required. They can be expressed in a simple form as:
%
\begin{align}
    \dot{\rho}_{r} + 4 H \rho_{r} =0,\label{sec2:eqn4.1} \\ 
 \dot{\rho}_{m} + 
  3 H (\rho_{m} + \Pi) =0.
  \label{sec2:eqn4}
\end{align}
Most existing works solve the corresponding evolution equations for particular values of the exponent $s$, as it is challenging to write them in the form of an autonomous system, which is necessary for applying dynamical system techniques. However, one interesting feature of both parametrizations is that the system maintains its autonomous form for any arbitrary value of $s$. Therefore, we can study the dynamical character of the fixed points and their corresponding stability properties of unified viscous models within this general setup. Before delving into that, we will discuss in the next section some physical conditions required to achieve accelerated expansion within this scenario.

\section{Condition for accelerated expansion from bulk viscosity}
\label{sec:conditions}

The condition for an accelerated expansion within this framework leads us to the question  of how much bulk viscosity the fluid must have to mimic the effect of the cosmological constant. This question can be answered by a simple examination of the condition for acceleration equation at very low redshift for the given effective pressure eqn. (\ref{sec2:eqn2}). This translates into the condition 
\begin{equation}
   w_{m}^{\rm eff}\equiv-\frac{3H\xi}{\rho_{m}}<-\frac{1}{3},
\end{equation}
which can be expressed, in turn, as
\begin{equation}
    \xi>\frac{H \Omega_{m}}{8\pi G_{N}}.
\end{equation}
This inequality holds only from a certain time at which the bulk viscosity coefficient overcomes the right-hand side, giving rise to the onset of the acceleration in the cosmic expansion. In practice, we evaluate the above expression at the present time $t_{0}$, whose inequality still holds. As reference values, we assume Planck 2018 results \cite{Planck:2018vyg}\footnote{We have simply estimated the matter density parameter from the baryon density parameter as follows: $\Omega_{m,0}=1-\Omega_{b}$, where we have neglected radiation for the sake of example. This relation is equivalent to requiring an effective energy density and pressure today that match respectively $\rho_{\rm m,0}\approx \rho_{\rm CDM,0} + \rho_{\rm \Lambda}$ and $p_{\Lambda}=-\rho_{\Lambda}\approx-3H\xi$, to be in agreement with late-time observations.}, leading to
\begin{equation}
    \xi (t=t_{0}) > 1.123\times 10^{8} \left(\frac{H_{0}} {67.4~\rm{km/s/Mpc}}\right) \left(\frac{ \Omega_{m,0}}{0.96}\right) \rm{Pa.s},\label{sec5:constraint1}
\end{equation}
which is fully independent of the assumed parameterization and consistent with early estimations (see e.g., \cite{PAVON1993261}). Nevertheless, this value is difficult to understand in terms of what we expect from typical values of bulk viscosity and fluid density. On one side, we have extreme density in astrophysical systems, such as neutron stars, which typically are of the order of $\rho\sim 10^{17}$ kg/m$^{3}$; and on the other side, we have very low cosmic density of dark matter on the order of $10^{-27}$ kg/m$^{3}$. This implies a difference of $10^{44}$ orders of magnitude between these extreme astrophysical and cosmological fluids. It is worth noting that the highest value for a constant bulk viscosity coefficient in neutron star merger simulations is estimated to be $\xi \sim 10^{29}$ Pa·s \cite{Most:2022yhe}, which is significantly larger than the lower limit (\ref{sec5:constraint1}) by about $10^{21}$ orders of magnitude. Given the huge differences in their respective densities, one could expect even more contrasting values of bulk viscosity. In fact, from what we know about typical fluids in terrestrial environments, bulk viscosity is present even in monoatomic gases with small values on the order of $\mathcal{O} (10^{-10})$ Pa·s \cite{sharma2023bulk}.

Despite not knowing the nature of the dark matter component, it is highly improbable, as discussed above, that some kind of field or particle would exhibit such huge values of bulk viscosity at very low present densities. Such large values of bulk viscosity, necessary for achieving accelerated expansion, would also dramatically impact the evolution of cosmic structures unless there is a physical mechanism that dissipates the viscosity over time during structure formation and makes it relevant only close to the present era, thereby triggering cosmic acceleration. However, this scenario would require fine-tuning in contrast to a smoother behavior for the bulk viscosity coefficient.

Let us consider models of the form $\xi \propto \rho_{m}^{s}$ with $s<0$, implying that the bulk viscosity dynamically diminishes towards the past as the energy density of dark matter increases. This counterintuitive behavior was necessitated in \cite{Bulk2,DiValentino:2021izs} to fit perturbative and background cosmological data. In other words, the unified viscous scenario suffers from the drawback of lacking a physical behavior for the bulk viscosity. Moreover, the dynamical system analysis of this viscous model indicates, as we shall see, that this inverse dependence on the matter density exacerbates our initial concern, as it prevents having a complete and consistent description of cosmic evolution\footnote{At the background level, it is well known that this model exhibits a behavior similar to the generalized Chaplygin gas model \cite{Bento:2002ps}.}.

The next question we want to address is whether current estimations of the bulk viscosity coefficient based on statistical analysis are consistent with the constraint (\ref{sec5:constraint1}). This concern arises because this information is not usually included in the priors on the model parameters. Parameter estimation procedure constrains, instead, the dimensionless quantity $\hat{\xi}_{0}$ which does depend on the assumed parameterization, such as models (\ref{sec2:eqn3}) and  (\ref{sec2:viscous_II}). Not surprising, this latter quantity can be linked to the lower limit we derived above. For instance, considering the model (\ref{sec2:viscous_II}), one can establish a more stringent condition involving both parameters $\hat{\xi}_{0}$ and $s$ to generate acceleration:
\begin{equation}
\hat{\xi}_{0}>\frac{\Omega_{m,0}^{1-s}}{9}.\label{sec5:constraint2}
\end{equation}
Considering the observational constraints on the model parameters for this model found in \cite{Yang:2019qza}, where $s=-0.557$, $\Omega_{m,0}=0.96$, and $\hat{\xi}_{0}=0.403$ \cite{Yang:2019qza}, we aim to verify whether these values align with the acceleration condition (\ref{sec5:constraint2}). This condition demands $\hat{\xi}_{0}>0.104$,\footnote{It is worth noting that in their study, they refer to $\hat{\xi}_{0}$ as $\beta$.} which validates the independent analysis based on statistical inferences but not resolve the discrepancies we discussed earlier. 

On the other hand, the inclusion of the cosmological constant can considerably lower the bound (\ref{sec5:constraint1}) because bulk viscosity turns out to be a subdominant effect in the cosmological dynamics. Some studies suggest, for example, $\xi \sim 10^{6}~\rm{Pa.s}$ \cite{valorxi1,valorxi2}, with small differences between them due to the assumed parametrization. 


Turning to the discussion of viscous unified scenarios, they must undergo scrutiny to ensure they provide a complete cosmological evolution, ensuring the cosmological viability of the model or indicating its failure. We will explore this delicate aspect in detail in the following sections.

\section{Dynamical system analysis for the viscous model I}
\label{sec:dynsys:symbolic}
Now we introduce dimensionless variables that define the phase space of the system. The inclusion of bulk viscosity in the form of Eqn.~(\ref{sec2:eqn3}) does not require an extra variable to close the system. This viscous fluid itself represents the dark sector of the Universe. Therefore, our variables correspond to the energy density parameters associated with each fluid component: 
%
\begin{align}
& \Omega_{r} \equiv \frac{8\pi G_{N} \rho_{r}}{3H^{2}}; \;\; \Omega_{m} \equiv \frac{8\pi G_{N} \rho_{m}}{3H^{2}}.
\label{sec3:eqn1}
\end{align}
Hence, the Friedmann constraint takes the simple form
\begin{equation}
\Omega_{r}+\Omega_{m}=1.\label{sec3:eqn2}
\end{equation}
The acceleration equation (\ref{sec2:eqn1}) involves an explicit dependence on the bulk viscosity coefficient, which is also manifested in the continuity equations (\ref{sec2:eqn4.1}) and (\ref{sec2:eqn4}).  The dynamical system associated with these phase space variables $(\Omega_{r}, \Omega_{m})$ can be written as an autonomous system in the following form:
\begin{align}
&\Omega_{r}^\prime=-\Omega_{r} (\Omega_{m} + 3 \hat{\xi}_{0} \Omega_{m}^{s} ), \;\; \quad \text{and} \quad
\Omega_{m}^\prime= (1-\Omega_{m}) (\Omega_{m} + 3 \hat{\xi}_{0} \Omega_{m}^{s} ), \label{sec3:eqn3}
\end{align}
where the prime denotes derivative with respect to $N\equiv\ln a$. These expressions explicitly show a dependence on the bulk viscosity, which can alternatively be interpreted as a non-linear interaction among both components. This feature leads to non-trivial effects on the background dynamics.
It it worth pointing out that the Friedmann constraint Eqn.~(\ref{sec3:eqn2}) could be used to reduce the dynamical system to just one evolution equation. However, we choose to keep it as a two-dimensional system to obtain a more intuitive representation of the evolution. The effective EoS parameter is defined as
\begin{equation}
w_{\rm eff}=-\frac{2}{3}\frac{H^\prime}{H}-1, \quad \text{with}\quad \frac{H^\prime}{H}=\frac{1}{2} (-3 + 3 \hat{\xi}_{0}  \Omega_{m}^{s} - \Omega_{r}).\label{sec3:eqn4}
\end{equation}
%
As we saw in the last section, one expects from a physical perspective that the viscous coefficient fulfills the condition $\hat{\xi}_{0}\sim\mathcal{O}(1)$, such that the viscosity leads to an accelerated expansion in the absence of the cosmological constant. Similarly, using Eqn.~(\ref{sec2:eqn1}), the deceleration parameter can be expressed as
\begin{equation}
q 	\doteq - \frac{\ddot{a} / a} {H^2} =  \frac{1}{2} \hspace{0.1 cm} (1+ \Omega_{r} ) - \frac{3}{2} \hspace{0.1 cm} \hat{\xi}_{0} \hspace{0.1 cm} \Omega_{m}^{s}.
\label{accel_exp}
\end{equation}
Having described the dynamical equations for this viscous scenario, we can proceed with a detailed analytical analysis of the asymptotic behavior of the background dynamics. The advantage of our approach lies in its validity for arbitrary values of the exponent $s$.

\begin{table*}[htp]
\centering  
\caption{Fixed points of the autonomous system described by Eqn.~(\ref{sec3:eqn3}) for different values of the bulk viscosity exponent $s$ along with the condition of existence of the fixed point. The main cosmological features of the model have been also included.}
\begin{ruledtabular}
\begin{tabular}{ccccccccccc}
Exponent& Point & $\Omega_{r}$ & $\Omega_{m}$ & $w_{\rm eff}$ & \text{Existence} & \text{Acceleration}
  \\ \hline
  \multirow{1}{*}{$s=-2$} &  
 $(\rm Aa)$ & $0$ & $1$ & $ -\hat{\xi}_{0} $ &  $\forall \hat{\xi}_{0}$ & $\text{Yes}$\\
  \\ \hline
  \multirow{3}{*}{$s=-\frac{3}{2}$} &  
 $(\rm Ba)$ & $0$ & $1$ & $\hat{\xi}_{0}$ &  $\forall \hat{\xi}_{0}$ & $\text{No}$\\
 & $(\rm Bb)$ & $0$ & $1$ & $-\hat{\xi}_{0}$ &  $\forall \hat{\xi}_{0}$ & $\text{Yes}$\\
 & $(\rm Bc)$&$1-(3\hat{\xi}_{0})^{2/5}$
 & $(3\hat{\xi}_{0})^{2/5}$ & $\frac{1}{3} \left(1 - 2 (3\hat{\xi}_{0})^{2/5} \right)$ & $\forall \hat{\xi}_{0}$ & $\text{No}$\\
 \\ \hline
  \multirow{1}{*}{$s=-1$} & $(\rm Ca)$ & $0$ & $1$ & $-\hat{\xi}_{0}$ &  $\forall \hat{\xi}_{0}$ & $\text{Yes}$\\
 \\ \hline
  \multirow{4}{*}{$s=-\frac{1}{2}$} &  
 $(\rm Da)$ & $0$ & $1$ & $-\hat{\xi}_{0}$ &  $\forall \hat{\xi}_{0}$ & $\text{Yes}$\\
 & $(\rm Db)$ & $0$ & $1$ & $\hat{\xi}_{0}$ &  $\forall \hat{\xi}_{0}$ & $\text{No}$\\
  & $(\rm Dc)$ &$1-(3\hat{\xi}_{0})^{2/3}$
 & $(3\hat{\xi}_{0})^{2/3}$ & $\frac{1}{3} \left(1 - 2 (3\hat{\xi}_{0})^{2} \right)$ & $\hat{\xi}_{0}>0$ & $\text{No}$\\
 \\ \hline
  \multirow{1}{*}{$s=0$} &  
 $(\rm Ea)$& $0$& $1$ & $-\hat{\xi}_{0}$ & $\forall \hat{\xi}_{0}$ & $\text{Yes}$\\ 
 \\ \hline
  \multirow{3}{*}{$s=\frac{1}{2}$} &  
 $(\rm Fa)$ &$0$ & $1$ & $\hat{\xi}_{0}$ & $\forall \hat{\xi}_{0}$ & $\text{No}$\\
 & $(\rm Fb)$  &$0$ & $1$ &$-\hat{\xi}_{0}$ & $\forall \hat{\xi}_{0}$ & $\text{Yes}$ \\
 & $(\rm Fc)$ & $1-(3\hat{\xi}_{0})^{2}$
 & $(3\hat{\xi}_{0})^{2}$ & $\frac{1}{3}$ & $\forall \hat{\xi}_{0}$ & $\text{No}$\\
  \\ \hline
  \multirow{2}{*}{$s=1$} &  
 $(\rm Ga)$ & $0$ & $1$ & $-\hat{\xi}_{0}$ & $\forall \hat{\xi}_{0}$ & $\text{Yes}$\\
 & $(\rm Gb)$ & $1$ & $0$ & $\frac{1}{3}$ & $\forall \hat{\xi}_{0}$ & $\text{No}$\\
  \\ \hline
  \multirow{4}{*}{$s=\frac{3}{2}$} &  
 $(\rm Ha)$ & $0$ & $1$ & -$\hat{\xi}_{0}$ & $\forall \hat{\xi}_{0}$ & $\text{Yes}$\\
 & $(\rm Hb)$ & $0$ & $1$ & $\hat{\xi}_{0}$ & $ \forall \hat{\xi}_{0}$ & $\text{No}$\\
 & $(\rm Hc)$ & $1$ & $0$ & $\frac{1}{3}$ & $\forall \hat{\xi}_{0}$ & $\text{No}$\\
  & $(\rm Hd)$ & $1-(3\hat{\xi}_{0})^{-2}$
 & $(3\hat{\xi}_{0})^{-2}$ &  $\frac{1}{3} \left(1 - \frac{2}{(3\hat{\xi}_{0})^{2}} \right)$ & $\hat{\xi}_{0}\neq0$ & $\text{No}$\\
  \\ \hline
  \multirow{2}{*}{$s=2$} &  
 $(\rm Ia)$ & $0$ & $1$ & $-\hat{\xi}_{0}$ & $\forall \hat{\xi}_{0}$ & $\text{Yes}$\\
 & $(\rm Ib)$ & $1$ & $0$ & $\frac{1}{3}$ & $\forall \hat{\xi}_{0}$ & $\text{No}$\\
 \\
 \end{tabular}
\end{ruledtabular}\label{M1:critical points}
\end{table*}
\begin{table*}[htp]
\centering  
\caption{Eigenvalues and stability conditions for determining the dynamical character of fixed points meeting the criteria for a suitable dynamical evolution: from a radiation-dominated era to accelerated expansion.}
\begin{ruledtabular}
\begin{tabular}{ccccccccccc}
Exponent & Point & $\lambda_{1}$ & $\lambda_{2}$ & \text{Stability} \\ \hline
 \multirow{2}{*}{$s=1$} &
$(\rm Ga)$ & $-(1+3\hat{\xi}_{0})$ & $-(1+3\hat{\xi}_{0}) $ &  \text{Attractor} for $0<\hat{\xi}_{0}<1$ \\
 & $(\rm Gb)$ & $0$ & $1$ & $\text{Marginal/Reppeler} \;\text{if}\; \hat{\xi}_{0}>0$\\
\\ \hline
 \multirow{3}{*}{$s=\frac{3}{2}$} &
$(\rm Ha)$ & $-(1+3\hat{\xi}_{0})$ & $-(1+3\hat{\xi}_{0})$ & $\text{Attractor} \;\text{for}\; \hat{\xi}_{0}>1/3$\\
 & $(\rm Hc)$ & $1$ & $0$ &$\text{Repeller/Marginal} \; \forall\hat{\xi}_{0}>0$\\
& $(\rm Hd)$ & $-2/(3\hat{\xi}_{0})^2$ & $ \frac{1}{2} \left(5-1/(\hat{\xi}_{0})^2)\right)$ & $\text{Attractor/Repeller} \;\text{for}\; \hat{\xi}_{0} > 0.447$ \\ 
\\ \hline
 \multirow{2}{*}{$s=2$} &
$(\rm Ia)$ & $-(1+3\hat{\xi}_{0})$ & $-(1+3\hat{\xi}_{0})$ & $\text{Attractor} \;\text{for}\; \hat{\xi}_{0} > 1/3$\\
 & $(\rm Ib)$ & $0$ & $1$ & $\text{Marginal/Repeller} \; \forall\hat{\xi}_{0}>0$\\
\\
\end{tabular}
\end{ruledtabular}\label{M1:eigenvalues}
\end{table*}
%

%
%
\subsection{Analytical study of the fixed points structure for arbitrary bulk viscosity exponents}
\label{sec:dynsys:analytical}
We first calculate the fixed points of the dynamical system (\ref{sec3:eqn3}), followed by a linear stability analysis. Beyond the linear regime, we will apply the criterion of the general stability theorem set by Malkin \cite{Malkin_52}. To simplify the notation, we introduce the variables $X = \Omega_{r}$ and $Y = \Omega_{m}$. The fixed points fall generally into three solutions:
\begin{itemize}

\item  \textit{Type $I$}:   

\begin{equation}
X_I = 0, \quad Y_I = 1, \quad \text{with} \quad w_{eff} = -(1^s) \hat{\xi}_{0} \quad \text{and} \quad q = \frac{1} {2} - (1^s) \frac{3\hat{\xi}_{0}} {2}. 
\label{GFP_I} 
\end{equation}
This one-parameter family of fixed points represents physically viscous dark matter domination regardless of the values of $s$ and $\hat{\xi}_{0}$.\footnote{We remind the reader that here we refer to viscous dark matter as a single unified description of the dark sector of the universe.} They may lead to a cosmic acceleration if $\hat{\xi}_{0} > 1/3 $, considering the positive branch in the involved root, whereas for $\hat{\xi}_{0} \ll 1$ on the negative branch, the fluid behaves almost like standard dark matter (dust) with a very small dissipative pressure.\\

\item  \textit{Type $II$ (for $s>0$)}:

\begin{equation}
Y_{II} = 0, \quad X_{II} = 1, \quad \text{with} \quad w_{eff} = \frac{1}{3} \quad \text{and} \quad q = 1.
\label{GFP_II}
\end{equation}
This general fixed point represents a radiation fluid with standard EoS and decelerated expansion.\\

\item \textit{Type $III$ (for $s\neq 1 $)}:

\begin{equation}
\quad X_{III} = 1- Y_{III}, \quad Y_{III} = (-3 \hat{\xi}_{0})^{ 1 / (1-s)}, \quad \text{with} \quad w_{eff} = \frac{1}{3} X - \hat{\xi}_{0} Y^{s} \quad \text{and} \quad q =  \frac{1}{2} \left( 1+X\right) - \frac{3}{2} \hat{\xi}_{0} Y^{s}.
\label{GFP_III}
\end{equation} 
To ensure that $Y_{III} \in \mathbb{R}_{+} $, the exponent $s$ must be a rational number of the form $s = (1+2n)/2m$, where $n$ and $m$ are arbitrary integers.
This restriction results, respectively, in the following effective EoS and deceleration parameters:
\begin{equation}
w_{eff} = \frac{1}{3} \left( 1 - 2 (3 \hat{\xi}_{0})^{2m/(2(m-n)-1)} \right) \quad \text{and} \quad q =  1 - (3 \hat{\xi}_{0})^{2m/(2(m-n)-1)},
\label{GFP_IIIb}
\end{equation} 
This fixed point describes a universe with both radiation and dark matter components. Therefore, the dynamical character of this point can change depending on the values of the model parameters. In the limit as $\hat{\xi}_{0} \to 0$, we recover the solution for radiation domination (\ref{GFP_II}). This indicates that the presence of dark matter is fully supported by the bulk viscosity pressure. In this sense, this fixed point can lead to accelerated expansion on the condition that $\hat{\xi}_{0} > 1/3$, similar to the previous Type $I$ case.
\end{itemize}
After describing the conditions for the existence of the fixed points, which can exist simultaneously or not depending on the value of $s$, we proceed to illustrate the main features of these solutions for specific values of the exponent $s$. It is important to note that within our setup, any unified viscous model must either account for pure radiation domination (type $II$) or include radiation with a small contribution from bulk viscosity dark matter (type $III$). We provide examples that cover either one, two, or all three types of fixed points. These are detailed in Table \ref{M1:critical points}, along with some cosmological characteristics, as described below:\\

{i) $s = -2$. This value allows for the existence of a unique fixed point, which belongs to type $I$. Specifically, it takes the following form:
%
\begin{equation}
X_I= 0, \quad Y_I = 1 \quad \text{with} \quad w_{eff} = -\hat{\xi}_{0}, \quad \text{and} \quad q = \frac{1}{2} - \frac{3}{2} \hat{\xi}_{0}.
\end{equation}
This fixed point corresponds to a viscous DM component, denoted as point (Aa) in Table {I}, and can successfully drive  the current accelerated expansion provided $\hat{\xi}_{0}>1/3$. However, it fails to account for radiation domination and must be discarded as  possible unified dark model.\\

ii) $s = -3/2$. In this case, pure radiation domination (type $II$) is not possible. However, fixed points of type $I$ remain, labeled as (Ba) and (Bb), respectively. These can be summarized as
\begin{equation}
X_I= 0, \quad Y_I = 1, \quad \text{with} \quad w_{eff} = \mp \hat{\xi}_{0}, \quad \text{and} \quad q =  \frac{1}{2} \mp \frac{3}{2} \hat{\xi}_{0}.
\end{equation}
Finally, the fixed point of type $III$ (see Eqn. (\ref{GFP_III})) yields:
\begin{equation}
X_{III} = 1 - (3 \hat{\xi}_{0})^{2/5}, \quad Y_{III} = (3\hat{\xi}_{0})^{2/5}, \quad \text{with} \quad w_{eff} = \frac{1}{3} \left(1 - 2 (3\hat{\xi}_{0})^{2/5} \right) \quad \text{and} \quad q = 1 - (3\hat{\xi}_{0})^{2/5}. 
\end{equation}
This is denoted as (Bc) in Table I. Solutions of types $II$ and $III$ cannot coexist simultaneously under the physical requirement of complete cosmological evolution. If $\hat{\xi}_{0}>1/3$, leading to accelerated expansion, it disrupts the radiation era, rendering this solution unsuitable for a unified scenario.\\

iii) $s = -1$. In this case, only the fixed point of type I ((Ca) in the table) exists, describing viscous dark matter with the potential to generate accelerated expansion provided that $\hat{\xi}_{0} > 1/3$. This is characterized by
\begin{equation}
X_I= 0, \quad Y_I = 1, \quad \text{with} \quad w_{eff} = - \hat{\xi}_{0}, \quad \text{and} \quad q = \frac{1}{2} - \frac{3}{2} \hat{\xi}_{0}.
\end{equation}
However, it fails to describe the radiation era and must be discarded.\\

iv) $s = -1/2$. This value results in 
fixed points of type $I$, given by:
\begin{equation}
X_I= 0, \quad Y_I = 1, \quad \text{with} \quad w_{eff} = \mp \hat{\xi}_{0}, \quad \text{and} \quad q = \frac{1}{2} \mp \frac{3}{2} \hat{\xi}_{0},
\end{equation}
which corresponds to (Da)-(Db) in Table {I}. It also allows for a fixed point of type $III$, labeled as (Dc) in Table I. 
\begin{equation}
X_{III} = 1 - (3 \hat{\xi}_{0})^{2/3}, \quad Y_{III} = (3\hat{\xi}_{0})^{2/3},  \quad \text{with} \quad w_{eff} = \frac{1}{3} \left(1 - 2 (3\hat{\xi}_{0})^{2} \right), \quad \text{and} \quad q = 1 - (3\hat{\xi}_{0})^{2}. 
\end{equation}
It is important to notice that the fixed point of type $I$, corresponding to a viscous DM type, leads to accelerated expansion for $ \hat{\xi}_{0}>1/3$. However, for such values, $X_{III}$ becomes negative, which is physically unacceptable. Therefore, one can have either radiation domination for $ \hat{\xi}_{0}\ll 1$ or accelerated expansion for $ \hat{\xi}_{0}>1/3$, but unfortunately not both periods. As a consequence, this viscous model must also be discarded.\\

v) $s = 0$. This particular value corresponds to a bulk viscosity coefficient of the form $\xi \propto H$, commonly used in unified viscous models. In this scenario, there exists only one fixed point of type $I$, which describes dissipative dark matter. This is given by
\begin{equation}
X_I= 0, \quad Y_I = 1, \quad \text{with} \quad w_{eff} = - \hat{\xi}_{0}, \quad \text{and} \quad q = \frac{1}{2} - \frac{3}{2} \hat{\xi}_{0}.
\end{equation}
This fixed point, denoted as (Ea), leads to accelerated expansion for $\hat{\xi}_{0}>1/3$, but it fails to describe the radiation era of the universe. Therefore, it must be discarded. This is in line with \cite{Avelino:2013mua}.\\

vi) $s = 1/2$. This value corresponds to the viscous model $\hat{\xi} \propto \rho_{m}^{1/2}$, which has been extensively used in the literature. However, this particular case must be approached differently within our framework. While it seemingly exhibits a fixed point of type $II$ $(X_{II}=1, Y_{II}=0)$, it fails to satisfy the condition of Malkin's theorem Eqn.~(\ref{malkin}). This argument will be developed in more detail when studying the stability conditions. Instead, it presents two fixed points of type $I$, labeled as (Fa)-(Fb):
%
\begin{equation}
X_I= 0, \quad Y_I = 1, \quad \text{with} \quad w_{eff} = \mp \hat{\xi}_{0}, \quad \text{and} \quad q = \frac{1}{2} \mp \frac{3}{2} \hat{\xi}_{0}.
\end{equation}
Additionally, it exhibits one fixed point of type $II$ (see Eqn. (\ref{GFP_II})):\\

Type $II$ :
\begin{equation}
\quad X_{II} = 1, \quad Y_{II} = 0, \quad \text{with} \quad w_{eff} = \frac{1} {3}  \quad \text{and} \quad q = 1 .
\label{s_one_half_II}
\end{equation}
This fixed point describes a standard radiation dominated era, with the known EoS coefficient $\omega_{eff} = 1/3.$ \\

Type $III$:
\begin{equation}
X_{III} = 1 - (3 \hat{\xi}_{0})^2, \quad Y_{III} = (3\hat{\xi}_{0})^2, \quad \text{with} \quad w_{eff} =  \frac{1}{3} \left(1 - 2 (3\hat{\xi}_{0})^2 \right) \quad \text{and} \quad q = 1 - (3\hat{\xi}_{0})^2. \label{shalf_III}
\end{equation}
This latter is denoted as (Fc) in Table I. While this scenario appears to present a complete cosmological evolution—featuring a radiation domination era with non-vanishing viscous dark matter and an accelerated expansion driven by bulk viscosity pressure—this fixed point must be disregarded, as $X_{III}$ becomes negative for $\hat{\xi}_{0}>1/3$. Relaxing this condition inevitably compromises the behavior of dark energy at late times. This scenario faces similar issues as previous cases and, surprisingly, should not be considered as a unified description of the dark sector. \\

vii) $s=1$. This particular case must be handle separately, as the dynamical system simplifies its analytical dependence, and moreover, has no fixed points of type $III$. Accordingly, the fixed point of type $I$ is
\begin{equation}
X_I= 0, \quad Y_I = 1, \quad \text{with} \quad w_{eff} = - \hat{\xi}_{0} \quad \text{and} \quad q = \frac{1}{2} - \frac{3}{2} \hat{\xi}_{0}, 
\end{equation}
which corresponds to (Ga) in Table I.  A fixed point of type $II$ is also present
\begin{equation}
X_{II}= 0, \quad Y_{II} = 1, \quad \text{with} \quad w_{eff} = 1/3, \quad \text{and} \quad q = 1,  
\end{equation}
which corresponds to (Gb) in Table I. This parametrization correctly captures both radiation and matter-dominated periods, and the viscous pressure could drive accelerated expansion for $\hat{\xi}_{0}>1/3$.\\ 

In addition, for this case it is possible to integrate analytically the dynamical system. The behavior for the normalized DM energy density as a function of the scale parameter $a$ is given by
\begin{equation}
\Omega_m (a) = \frac{a^{1 + 3 \hat{\xi}_{0}}}{1 + a^{1 + 3 \hat{\xi}_{0}}}, 
\label{analytic_Y_s_1_diss_I}
\end{equation}
which fulfills the suited conditions to vanishes as $a\rightarrow{0}$, or equivalently, $\Omega_m$ vanishes asymptotically in the radiation dominated epoch. This asymptotic behavior corresponds to the fixed point of type {II}. It is worthwhile pointing out that deep in this era and in the zero viscosity limit $ \hat{\xi_{0}}$, the dark matter normalized energy density given by the above equation behaves as $\Omega_m (a) \propto a $. But, in this asymptotic region the square of the Hubble parameter goes proportional to the radiation energy density (see Eqn.(\ref{sec2:eqn0})), and as a consequence, the dark matter energy density behaves as a power law of the scale parameter with the well-known exponent, i.e. $\rho_m (a) \propto a^{-3}$. This result represents a consistency check of the validity of expression Eqn.(\ref{analytic_Y_s_1_diss_I}), obtained by a direct integration of the dynamical system governing the cosmological evolution.\\  
Similarly, for the DM-dominated era $a\rightarrow{\infty}$, $Y \rightarrow{1} $, whose asymptotic behavior is described by the fixed point of type {I}. As a consequence, the effective DM pressure of Eqn.~(\ref{sec2:eqn2}) can be expressed in the form
\begin{equation}
\Pi (a) = - \hat{\xi}_{0} ~ \rho_{m} = - \frac{3H^{2} \hat{\xi_{0}}}{8\pi G_{N} } \frac{a^{1 + 3 \hat{\xi}_{0}}}{1 + a^{1 + 3 \hat{\xi}_{0}}} . 
\label{analytic_Pressure_s_1_diss_I}
\end{equation}
As $H^{2}$ is non-negative quantity and $\hat{\xi_{0}} > 0$, it follows that the negative effective pressure drives the universe to an accelerated expansion when the scale parameter $a$ grows, as expected by the cosmological observation data. On the contrary, it becomes negligible as $a$ becomes relatively small, compatible with the decelerated expansion of the radiation dominated epoch, including the Big Bang singularity ($a \rightarrow{0} $) . \\

viii) $s = 3/2$. This case exhibits similar features as the $s=1/2$ case with one key distinction we discuss later. There are four fixed points.  Two of the form type $I$
\begin{equation}
X_I= 0, \quad Y_I = 1, \quad \text{with} \quad w_{eff} = \mp \hat{\xi}_{0}, \quad \text{and} \quad q = \frac{1}{2} \mp \frac{3}{2} \hat{\xi}_{0}
\end{equation}
which correspond to (Ha)-(Hb). In this scenario, accelerated expansion can take place, as usual, only in the positive branch of the square root of the above expressions for $\hat{\xi}_{0}>1/3$. Another fixed point reads
\begin{equation}
X_{II}= 1, \quad Y_{II} = 0, \quad \text{with} \quad w_{eff} = 1/3, \quad \text{and} \quad q = 1
\end{equation}
which corresponds to (Hc), and represents a radiation fluid. The last fixed point is of type $III$ and yields: 
\begin{equation}
\quad X_{III} = 1- Y_{III}, \quad Y_{III} = (3 \hat{\xi}_{0})^{-2}, \quad \text{with} \quad w_{eff} = \frac{1}{3} \left(1 - \frac{2}{(3\hat{\xi}_{0})^{2}} \right) \quad \text{and} \quad q = 1 - \frac{1}{(3\hat{\xi}_{0})^{2}}.
\label{s_III}
\end{equation}
This is marked as (Hd) in Table I. We have also considered the positive branch of the square root of the unity, consistent with the choice made for the fixed point of type $I$. 
As anticipated, we can see a key difference with respect the case $s=1/2$ in the fixed point of type $III$ (compare Eqns.~(\ref{shalf_III}) and (\ref{s_III})). For $\hat{\xi}_{0}>1/3$, this fixed point exists thanks to the negative square in $Y_{III}$, which prevents this quantity to become larger than one, as happened in the $s=1/2$ case, leading to a positive value for $X_{III}$. 
Interestingly, this fixed point describes a combined radiation-viscous matter fluid with a generalized EoS that encapsulates the effects mentioned, but it is not compatible with an accelerated expansion. \\

ix) $s=2$. This fixed point shares the same attributes as the $s=1$ case, so we will skip this discussion. However, it is worth noting that the stability properties may potentially differ.\\

We conclude this section by discussing the implications of considering arbitrary exponents in the fixed points analysis of the dynamical system described by Eqn.~(\ref{sec3:eqn3}). When dealing with non-positive integer exponents, denoted as $s = -n$ with $n= 0,1,2,3 ...$, we observe that only fixed points of type $I$ are present. This is because the condition $X=1$ and $Y=0$ to obtain radiation domination fails to yield a stationary point for Eqn.~(\ref{sec3:eqn3}). There remains the option of having fixed points of type $III$. However, for a general negative irrational exponent value $s$, the factor $(-1)^{ 1 / (1 - s)}$ appearing in Eqn.~(\ref{GFP_III}) introduces an imaginary component given by the expression $ \sin ( \pi (2n+1) / (1-s)) $ with $m \in \mathbb{Z}$, which is physically unacceptable. We conclude that for negative irrational $s$-values, the fixed points of Type $III$ must be discarded. Consequently, in a vicinity of the fixed point associated with $s=-1/2$, whose exponent already fails to describe the radiation era of the universe, for instance $s=-1/2+\epsilon$ with $\epsilon \ll 1$, there will not be a value yielding a complete description of the universe's evolution.
As a main result of this part, it is imperative to exclude such exponents from consideration in phenomenological viscous unified models, as they do not lead to physically meaningful solutions capable of describing the full dynamics of the universe.

\subsection{Stability of the fixed points}
In this subsection, we investigate the stability properties of the fixed points identified in the previous section. Our objective is to ascertain whether these fixed points accurately portray the evolution of the universe. To achieve this, we employ linear stability analysis, a method well-suited for examining the behavior of the system in the vicinity of the critical points defined by the dynamical system (\ref{sec3:eqn3}). We compute the eigenvalues of the system derived from the Jacobian matrix, obtained from the first-order derivatives of the functions appearing on the right-hand side of Eqn. (\ref{sec3:eqn3}):
\begin{align}
F_1(X,Y)=-X (Y + 3 \hat{\xi}_{0} Y^{s} ), \quad F_2(X,Y)= (1-Y) (Y + 3 \hat{\xi}_{0} Y^{s} ).  
\end{align}
From here, we obtain
\begin{align}
F_{1,X}&=-(Y +  3 \hat{\xi}_{0} Y^{s}), \hspace{1.2 cm} F_{2,X} = 0,
\label{lin_deriv_1}\\
F_{1,Y}&= -X (1 + 3s \; \hat{\xi}_{0} Y^{s-1}) , \quad F_{2,Y} = 1-2Y + 3\hat{\xi}_{0} (s\; Y^{s-1}-(1+s) Y^{s}),  
\label{lin_deriv_2}
\end{align}
where the comma indicates derivative with respect to the variable. We proceed then to compute the characteristic equation:
\begin{equation}
0 = \det (\mu I - a_{ij}), \quad \text{where} \quad a_{ij} = F_{i,j} .    
\end{equation}
By solving this characteristic equation, we can obtain the eigenvalues for each fixed point. Subsequently, we can determine whether these eigenvalues correspond to attractors, repellers, or saddle points, which in turn define the dynamical character of the fixed points. 
Regarding the applicability of linear stability theory beyond this regime, it is essential to address a technical point. Malkin's nonlinear stability theorem \cite{Malkin_52} asserts that the conclusions drawn from linear stability analysis concerning a fixed point $(X_*, Y_*)$ remain valid beyond the linear approximation if the right-hand side functions of the dynamical system defined by Eqn. (\ref{sec3:eqn3}), denoted as $F_i(X, Y)$, satisfy the following condition in a sufficiently small neighborhood of the fixed point:
\begin{equation}
\left|F_i (X, Y) \right| \leq \mathcal{N} \left(X^{2} + Y^{2} \right)^{1/2 + \alpha}.
\label{malkin}
\end{equation}
where $\left|F_i \right|$ stands for the absolute value of the function $F_i$, and $\mathcal{N}$ and $\alpha$ are positive constants, together with the additional condition on the real parts of the eigenvalues of the characteristic equation associated with the Jacobian matrix of being negative (stable fixed point), or being at least one of them positive (unstable fixed point). This mathematical aspect is crucial because in the open interval $0<s<1$, the above condition is violated by $F_2$ for fixed points of type {II}. Consequently, the stability analysis demands a higher-order examination. Now, we delve into the stability properties of suitable fixed points associated with those exponent values for which both types of fixed points exist—radiation and viscous matter—and additionally lead to an accelerated expansion. \\

i) For $s=1/2$, in principle, there are two fixed points that fulfill the aforementioned conditions.
Nevertheless, the dynamical system has ill-defined derivatives at the fixed point $(X=1,Y=0)$ as it can be seen from Eqn.~(\ref{lin_deriv_2}). In addition, the condition (\ref{malkin}) for the validity of the linear stability analysis beyond the linear approximation is violated, and therefore, we have to deal with this technical issue by performing instead a numerical analysis.  


%
%
The stability properties of the fixed points for this case are as follows:\\

Type $I$ :   

\begin{equation}
X_I = 0, \quad Y_I = 1, \quad \text{with} \quad w_{eff} = \mp \hat{\xi}_{0} \quad \text{and} \quad q = \frac{1}{2} \mp \frac{3\hat{\xi}_{0}}{2}.
\end{equation}
As we can see, this fixed point corresponds to viscous dark matter, leading to accelerated expansion under the aforementioned condition ($\hat{\xi}_{0} > 1/3$). The associated eigenvalues are degenerated and given by

\begin{equation}
\mu_{1,2} = -( 1 \pm 3 \hat{\xi}_{0}).  
\end{equation}
These eigenvalues describe an attractor fixed point, provided one chooses the positive branch of the square root. This condition is consistent with the above constraint ($\hat{\xi}_{0} > 1/3$), such that viscosity drives the universe to an accelerated expansion in the dark-matter dominated era. \\

Type $II$ :
\begin{equation}
\quad X_{II} = 1, \quad Y_{II} = 0, \quad \text{with} \quad w_{eff} = \frac{1} {3}  \quad \text{and} \quad q = 1 .
\label{s_one_half_II}
\end{equation}
This fixed point exists, as it is an stationary point of the dynamical system for $s = 1/2$, and describes a standard radiation dominated era. Nevertheless, as it was explain above, its stability properties go beyond the linear stability analysis. This aspect will be addressed by numerical analysis.\\

Type $III$ :
\begin{equation}
X_{III} = 1 - (3 \hat{\xi}_{0})^2, \quad Y_{III} = (3\hat{\xi}_{0})^2, \quad \text{with} \quad w_{eff} =  \frac{1}{3} \left(1 - 2 (3\hat{\xi}_{0})^2 \right) \quad \text{and} \quad q = 1 - (3\hat{\xi}_{0})^2. 
\label{s_one_half_III}
\end{equation}

This type of fixed point describes a combined radiation-matter fluid with a non-standard EoS parameter due to the presence of bulk viscosity. However, when the condition for accelerated expansion is fulfilled, the energy density parameter for radiation becomes negative, which is physically unsatisfactory\footnote{This type of critical point is also absent in a more general scenario where viscous dark matter and dark energy fluid are allowed to interact, and the bulk viscosity is a function of the total energy density \cite{Avelino:2013wea}.}. Therefore, fixed points of type $III$ must be discarded as it is not compatible with the existence of the other fixed points.\\

ii) For $s=1$, there are two fixed points that fulfill the conditions necessary to describe the complete evolution of the universe. For the first fixed point of type $I$, where $X_I=0$ and $Y_I=1$, and considering Eqns.~(\ref{lin_deriv_1}) and (\ref{lin_deriv_2}), we observe that the two eigenvalues are degenerate:
\begin{equation}
\mu_{1,2} = -(1+3\hat{\xi}_{0}) \quad \text{with} \quad w_{eff} = - \hat{\xi}_{0} \quad \text{and} \quad q = \frac{1}{2} - \frac{3}{2} \hat{\xi}_{0}. 
\label{s_one_I}
\end{equation}
This point corresponds to (Ga) in Table \ref{M1:eigenvalues}. As both eigenvalues are negative, the associated fixed point corresponds to an attractor, and represents a viscous matter capable of driving the accelerated expansion, provided that  $\hat{\xi}_{0} > 1/3$. For the second fixed point of type $II$, $X_{II}=1$, $Y_{II}=0$, the eigenvalues, displayed as (Gb) in Table \ref{M1:eigenvalues}, are given by
\begin{equation}
\mu_1 = 1 +3 \hat{\xi}_{0}, \quad \mu_2 = 0 \quad \text{with} \quad w_{eff} = \frac{1}{3} \quad \text{and} \quad q =  1.  
\end{equation}
These eigenvalues describe a repeller/marginal fixed point, corresponding to standard radiation dominated era. \\

iii) For $s=3/2$ there are in principle three fixed points, with two of them fulfilling the required conditions to describe the complete evolution of the universe. For the first fixed point of type $I$, $X_I=0$, $Y_I=1$, the two eigenvalues are degenerated as in the $s=1/2$ case:
\begin{equation}
\mu_{1,2} = -(1+3\hat{\xi}_{0}) \quad \text{with} \quad w_{eff} = \mp \hat{\xi}_{0} \quad \text{and} \quad q = \frac{1}{2} \mp \frac{3}{2} \hat{\xi}_{0}. 
\label{s_one_I}
\end{equation}
As both eigenvalues are negative, the associated fixed point corresponds to an attractor, representing a viscous matter mimicking the effect of dark energy at late times. Again, the condition $\hat{\xi}_{0} > 1/3$ must be imposed to ensure accelerated expansion, along with the selection of the positive branch of the square root of 1.
For the second fixed point of type $II$, $X_{II}=1$, $Y_{II}=0$, the eigenvalues are:
\begin{equation}
\mu_1 = 1 , \quad \mu_2 = 0 \quad \text{with} \quad w_{eff} = \frac{1}{3} \quad \text{and} \quad q = 1.
\label{s_one_II}
\end{equation}
These eigenvalues describe a repeller/marginal fixed point, which describes standard radiation era. 
Finally, the eigenvalues associated with the fixed point of type $III$ are:
\begin{equation}
\mu_1 = \frac{-2} {(3\hat{\xi}_{0})^{2}} , \quad \mu_2 = \frac{1} {2}  \left(5- \frac{1}{\hat{\xi}_{0}^{2}} \right), \quad \text{with} \quad w_{eff} = \frac{1}{3} \left(1 - \frac{2}{(3\hat{\xi}_{0})^{2}} \right) \quad \text{and} \quad q = 1 - \frac{1}{(3\hat{\xi}_{0})^{2}}.
\label{s_one_half_III}
\end{equation}
This type of fixed point describes a combined radiation-matter fluid with a modified  effective EoS coefficient, when compared with a pure radiation fluid EoS (see Eqn.~(\ref{s_one_II})). Nevertheless, the physical constraint $Y_{III} \leq 1$, or equivalently $ \hat{\xi}_{0} \geq 1/3$, forbids this fixed point to describe an accelerated expansion. As a consequence, this fixed point must be a repeller, at least in one direction, to be in accordance to the universe dynamics. This leads to the more demanding lower bound condition $\hat{\xi}_{0} > 0.447$, ensuring that $\mu_2 > 0$. \\

iv) For $s=2$, there are two fixed points fulfilling the required conditions to describe a suitable dynamics. For the first fixed point of type $I$, $X_I=0$, $Y_I=1$, the two eigenvalues are degenerated:

\begin{equation}
\mu_{1,2} = -(1+3\hat{\xi}_{0}) \quad \text{with} \quad w_{eff} = - \hat{\xi}_{0} \quad \text{and} \quad q = \frac{1}{2} - \frac{3}{2} \hat{\xi}_{0}. 
\label{s_two_I}
\end{equation}

As they are both negative, the associated fixed point corresponds to an attractor, describing a viscous matter driving the accelerated expansion if  $\hat{\xi}_{0} > 1/3$. As to the second fixed point of type $II$, $X_{II}=1$, $Y_{II}=0$, the eigenvalues are

\begin{equation}
\mu_1 = 1 , \quad \mu_2 = 0 \quad \text{with} \quad w_{eff} = \frac{1}{3} \quad \text{and} \quad q = 1.  \end{equation}

These eigenvalues describe a repeller/marginal fixed point, representing standard radiation. \\

Based on the stability analysis, we conclude that the cases $s=1$ and $s=3/2$ are suitable scenarios for unified viscous models. The case $s=1$ can straightforwardly be extended to all non-negative integers $s = n$, where $n \in \mathbb{N}$. Rational exponents such as $(2n+1)/2n$, which cover the former case iii), are also eligible for modeling viscosity within the DM sector. These results are summarized in Table \ref{M1:eigenvalues}.

\section{Dynamical system analysis for the viscous model II}
\label{sec:dynsys:viscous_II}

Now, we compute the fixed points associated with the second type of parametrization of the bulk viscosity coefficient $\xi^{II} \propto \rho_{m}^{s}$ (Eqn.~(\ref{sec2:viscous_II})), followed by their stability properties. This model has been extensively studied in the literature in different cosmological scenarios. In particular, we want to draw attention to the pioneering work \cite{Colistete:2007xi}, where, to the best of our knowledge, it was first noticed that for $s>1/2$, the fixed point type $I$, describing viscous matter domination, corresponds to a saddle point, while for $s<1/2$, it represents an attractor de-Sitter solution. In the present viscous model, we must retain this feature along with the existence of a radiation stage to provide a realistic cosmological model. Therefore, in the following analysis, particular attention must be paid to the case $s<1/2$. We have also verified that our analysis fully agrees with the results of \cite{Colistete:2007xi}, as we have found the same critical points and stability properties\footnote{The curious reader can observe that in the phase space analysis of \cite{Colistete:2007xi}, the case $s=1/2$ cannot be covered because this makes the solution singular, which differs from our approach since our system is well-behaved for all non-negative values of $s$. Another key difference is the choice of the dynamical variables. Colistete et al. chose distinctively the Hubble parameter $H$, whose value extends to infinity in phase space. This is the reason why they have projected onto the Poincaré sphere. In contrast, our variable $\eta$ (see eqn. (\ref{eta})) is confined to the range $0<\eta<1$, despite being a function of the Hubble parameter, thus allowing us to represent the entire phase space from the original variables.}. However, the inclusion of a radiation component in this viscous model leads to a significant change in the main conclusion, as we uncover from our analysis.

Contrary to the previous model, the phase space for this viscous model is spanned by three instead of two variables: $(X,Y,\eta)$. The corresponding dynamical system turns out to be
\begin{align}
\nonumber
X^\prime   &=  X ( X - 1 - 3 ~ \eta ~ Y^{s} ), \\ 
Y^\prime   &= X Y + 3 ~ \eta ~ (1-Y) Y^{s}, \label{dyn_sys_viscous_II} \\ 
\eta^\prime&= (s - 1/2) ~ \eta ~ ( -3 - X + 3 ~\eta ~ Y^{s} ), \nonumber
\end{align}
where $X$ and $Y$ are the same physical quantities defined in Eqn.~(\ref{sec3:eqn1}), and the additional dimensionless variable $\eta$, which was necessary to close the system, is defined by
\begin{equation}
   \eta = \left(24 \pi G_{N} \right)^{s-1/2} \hat{ \xi_0} ~ H^{2s-1}.
\label{eta}    
\end{equation}
The effective EoS  and deceleration parameter are given, respectively, by 
\begin{equation}
w_{eff} = \frac {1}{3} X - \eta ~ Y^s, \quad \text{and} \quad q = \frac {1}{2} ~ (1 + X - 3 ~ \eta ~ Y^s ).
\label{w_eff_accel}
\end{equation}
We remind that this model coincides with the previous one for the case $s = 1/2$. We exclude then this case in the forthcoming analysis.

\subsection{Analytical study of the fixed points structure for arbitrary bulk viscosity exponents}
\label{sec:dynsys: viscous_II}
Following the same reasoning as in the former case, we briefly describe the main steps in the computation of the fixed points. For exponents $s\neq1/2$, the fixed points can be classified into two types:\\

Type $I$ ($\eta_I \neq 0$):   
\begin{equation}
s \in \mathbb{R} \setminus {\{1/2\}} : \quad \eta_I = 1, \quad X_I = 0, \quad Y_I = 1, \quad \text{with} \quad w_{eff} = -1 \quad \text{and} \quad q = -1. 
\label{GFP_I} 
\end{equation}
This fixed point is independent of $s$ and represents viscous matter fully dominating the universe. In the asymptotic limit, it describes a de Sitter-like solution.\\

Type $II$ ($\eta_{II} = 0$):
\begin{equation}
s \in \mathbb{R} \quad : (\eta_{II})^a = 0, \quad X_{II} = 0, \quad Y_{II} = 1, \quad \text{with} \quad w_{eff} = 0 \quad \text{and} \quad q = 1/2.
\label{GFP_II_a}
\end{equation}
This fixed point, labeled as $a$, represents a standard non-relativistic dark matter fluid, and leads to a decelerated expansion, and hence, it does not correspond to the known cosmological evolution. In addition, from its stability properties, this fixed point turns out to be an attractor/marginal point in the physical phase space spanned by $(\Omega_r, \Omega_m)$, which doesn't fit to the know universe dynamics either. As a consequence, this fixed point must be discarded. 

The second possible fixed point of this type, labeled as $b$, is the following:
\begin{equation}
s > 0 :\quad (\eta_{II})^b = 0, \quad X_{II} = 1, \quad Y_{II} = 0, \quad \text{with} \quad w_{eff} = 1/3, \quad \text{and} \quad q = 1 ,
\label{GFP_II_b}
\end{equation}
which exists for arbitrary positive $s$-values, and describes a standard radiation fluid.

It is worth pointing out that with the exception of the particular value $s=1/2$, there are no fixed points of type $III$ in this viscous model as in the former viscous model I, which could have described rational exponents. This occurs because it is not possible to find a fixed point that meets the physical constraints $0 < X_{III} < 1$, $Y_{III} \neq 0$, and simultaneously allows the cancellation of the right-hand side of Eqns.~(\ref{dyn_sys_viscous_II}). This complication arises due to the presence of another dynamical equation for the system related to $\eta$, which differs from the former model. Consequently, a solution of the form $Y_{III} = ( -3 ~ \eta_{III} )^{1 / (1-s)} $ for $\eta \neq 0$ cannot simultaneously satisfy the condition $\eta' = 0$.\footnote{In an effective theory of relativistic viscous fluids, viscosity has no effect on the behavior of radiation either, i.e., $w_{eff} = 1/3$ \cite{Bemfica:2022dnk}, consistent with standard cosmology. We conjecture that this aspect is not intrinsically related to a causal first-order theory of relativistic viscous fluid dynamics but may be a consequence of the functional form of the bulk viscosity.}

%
A more significant implication arising from the structure of the fixed points is the unequivocal exclusion of negative exponents. These exponents are automatically deemed invalid, as they fail to yield fixed points associated with a radiation-dominated era. Hence, viscous models with $s<0$ must be ruled out as they cannot describe a complete cosmological evolution.
In this sense, it is pertinent to explicitly address the specific case of $s=-1/2$, as this exponent value corresponds to a preferred value in the best-fit analysis of extensive cosmological datasets \cite{Yang:2019qza}, which also encompassed gravitational wave standard sirens \cite{Yang:2023qqz}. Unfortunately, the constraint $s<0$ has been overlooked in those analysis as a (physical) prior information, leading to a best-fit values of $s$ falling outside the acceptable range. 
To discard any potential ambiguity in the definition of dynamical variables that might overcloud physical solutions, we introduce the variable change $Z = Y^{-1/2}$, leading to the following dynamical system: 
\begin{align}
\nonumber
&X^\prime= X ~ ( X - 1 - 3 \eta ~ Z ), \\
&Z^\prime= \frac{1}{2} \left( - X + 3\eta ~ (1 - Z^2 ) ~ Z \right)  Z , \label{sec4:eqn7}\\ \nonumber
&\eta^\prime= \eta  \left( -3 - X + 3\eta ~ Z \right) .
\end{align} 
A simple examination reveals that a radiation-like solution of the general form $(X=1, Y= 0, \eta)$ is explicitly excluded as a fixed point, as was also the case in the original dynamical system. In fact, there is only one fixed point that satisfies the constraint $X + Z^{-2} = 1$ and the physical condition $\eta \geq 0$:
\begin{equation}
X_I = 0, \quad Z_I = 1, \quad \eta_{I} = 1 \quad \text{with} \quad w_{eff} = -1 \quad \text{and} \quad q = 1. 
\label{s_minus_one_half_I} 
\end{equation}
The above fixed point represents a viscous dark matter fluid, which describes an asymptotically de Sitter expansion. However, since this case does not include a radiation fixed point, it must be dismissed as a viable viscous unified model. 

As to the limit value $s=0$, which describes a constant bulk viscosity coefficient within this setup  (see Eqn.~(\ref{sec2:viscous_II}), the dynamical system reduces to 
\begin{align}
\nonumber
X^\prime   &=  X ( X - 1 - 3 ~ \eta ), \\ 
Y^\prime   &= X Y + 3 ~ \eta ~ (1-Y), \label{dyn_sys_viscous_II_s_zero} \\ 
\eta^\prime&= \frac {1}{2} ~ \eta ~ ( 3 + X - 3 ~\eta ). \nonumber
\end{align}
For the first fixed point, we have 
\begin{equation}
X_I = 0, \quad Y_I = 1, \quad \eta_{I} = 1 \quad \text{with} \quad w_{eff} = -1 \quad \text{and} \quad q = -1. 
\label{s_minus_one_half_I} 
\end{equation}
This solution represents a scenario where viscous dark matter dominates, leading to an expansion resembling that of de Sitter expansion. The second fixed point is
\begin{equation}
X_{II} = 1, \quad Y_I = 0, \quad (\eta_{I})^b = 0 \quad \text{with} \quad w_{eff} = 1/3 \quad \text{and} \quad q = 1. 
\label{s_minus_one_half_II} 
\end{equation}
This point characterizes a standard radiation fluid. Consequently, as we will explore in the next subsection on stability properties, this exponent satisfies the physical conditions necessary to depict a complete evolution of the universe, consistent with earlier findings \cite{Avelino:2013mua,Sasidharan:2015ihq}.

A final remark concerns the particular exponent $s=1/2$. This is a very special value for the dynamical equations as $\eta^\prime$ vanishes automatically (see Eqn.~(\ref{dyn_sys_viscous_II})). As a result, the evolution equations go exactly into the same dynamical system as the viscous model $I$, with $\eta$ going to $\hat{\xi}_{0}$ (see Eqn.~(\ref{eta})). Thus, both models are fully equivalent for $s=1/2$, and the previous analysis conducted within the viscous model $I$ remains applicable to the present model $II$.\\
%

\subsection{Stability properties of the fixed points}
In this subsection, we investigate the stability properties of various fixed points that could potentially characterize the entire cosmic evolution. Employing a methodology akin to that used for the first viscous model, we will skip certain technical details. The functions featured on the right-hand side of Eqns.~(\ref{dyn_sys_viscous_II}) are:  
\begin{align}
\nonumber
F_1(X,Y, \eta )&=X ( X-1 - 3 \eta ~ Y^{s} ), \\ 
F_2(X,Y, \eta) &= XY + 3 \eta ~ (1-Y) Y^{s}, \label{stab_eta_viscous_II} \\ \nonumber
\Bar{F_3}(X,Y, \eta)&= (s-1/2) ~ \eta ~( -3-X + 3 \eta ~ Y^{s} ).  
\end{align}
As mentioned earlier, the case $s=1/2$ has been already analyzed, so it is omitted from the current analysis. Nevertheless, for arbitrary $ s \in \mathbb{R} \setminus {\{1/2\}}$, the stability properties of the fixed point of type I changes in the $\eta$-direction as $s$ crosses this particular value. In fact, for this fixed point Eqn. (\ref{GFP_I}) ($X_I = 0$, $Y_{I}= 1$, $\eta=1$), the eigenvalues are:
\begin{equation}
\mu_{1} = -4, \quad \mu_{2} = -3, \quad \text{and} \quad \mu_{3} = 3(s - 1/2) \quad \text{together with,} \quad w_{eff} = - 1, \quad \text{and} \quad  q = -1. 
\label{eigenvalues_s_viscous_II_I}
\end{equation}
This fixed point describes an attractor in the physical phase space directions $( \Omega_{r}, \Omega_{m})$, driving to an accelerated expansion due to the negative bulk viscosity pressure. Nevertheless, for the $\eta$-direction, $\mu_{3}$ represents an attractor if $s < 1/2$, while it becomes a repeller for $s > 1/2$.  Consequently, there exists a critical value around which a unified dark matter scenario emerges ($s < 1/2$), driving late-time accelerated expansion. This feature of the phase space was first observed in \cite{Colistete:2007xi} (see also \cite{Acquaviva:2018rqi,acquaviva2015nonlinear}). We have shown explicitly how this behavior arises in the appendix.\\

Now, for fixed points of type $II$ Eqn. (\ref{GFP_II_b}) ($X_{II} = 1$, $Y_{II}= 0$, $\eta ^{b}=0$), the eigenvalues are given by:
\begin{equation}
\mu_{1,2} = 1, \quad \text{and} \quad \mu_{3} = -4(s-1/2) \quad \text{together with,} \quad w_{eff} = 1/3, \quad \text{and} \quad  q = 1. 
\label{eigenvalues_s_n_viscous_II_II_b}
\end{equation}
Thus, this point describes a repeller in the physical phase space directions $( \Omega_{r}, \Omega_{m})$, indicating a pure radiation era. The repeller character of the $\eta$-direction for $s<1/2$ follows from its behavior: $\eta \propto H^{(2s-1)}$, whose value tends to zero as $H$ goes to infinity (see Eqn.~(\ref{eta})). This behavior holds in the radiation dominated era, approaching the Big Bang singularity. 

As an overall conclusion of the above analysis, the requirement of late-time accelerated expansion alongside the condition of radiation domination, i.e., $s > 0$, confines the physical region to $0 < s < 1/2$. Imposing the requirement of late-time accelerated expansion alongside the condition of radiation domination, i.e., $s > 0$, confines the physical region to $0 < s < 1/2$. This interval must be incorporated as a prior in statistical analyses, contrasting with the unjustified assumption made in \cite{Yang:2019qza,Yang:2023qqz}.\\

Finally, for the limit value $s=0$, which corresponds to a constant viscous coefficient, the eigenvalues are:\\

For fixed points of type {I}: ($X_I = 0$, $Y_{I}= 1$, $\eta=1$)
\begin{equation}
\mu_{1} = -4, \quad \mu_{2} = -3, \quad \text{and} \quad \mu_{3} = -3/2 \quad \text{together with,} \quad w_{eff} = - 1, \quad \text{and} \quad  q = -1,
\label{eigenvalues_s_n_viscous_II_I}
\end{equation}
which describes an attractor in the physical phase space directions $( \Omega_{r}, \Omega_{m})$. The attractor character of the $\eta$-direction follows from the functional form $\eta \propto H^{-1}$, which goes to a constant value for an asymptotically de Sitter solution described by a fixed point of type $I$, and prevents $\eta$ to deviate from its constant value.\\

For fixed points of type $II$: ($X_{II} = 1$, $Y_{II}= 0$, $\eta ^{b}=0$)
\begin{equation}
\mu_{1,2} = 1, \quad \text{and} \quad \mu_{3} = 2 \quad \text{together with,} \quad w_{eff} = 1/3, \quad \text{and} \quad  q = 1,
\label{eigenvalues_s_n_viscous_II_II_b_S_0}
\end{equation}
which corresponds to a repeller in the physical phase space directions $( \Omega_{r}, \Omega_{m})$, indicating a period dominated by standard radiation $w_{eff}=1/3$. The repeller character of the $\eta$-direction follows again from the behavior of $\eta \propto H^{-1}$,  which leads $\eta \rightarrow 0 $, as $H \rightarrow \infty$ in the very early universe limit.

For $s=0$, or equivalently constant bulk viscosity coefficient, the dynamical system can be analytically be integrated. Close enough to the asymptotic region $\eta = (24 \pi G_{N})^{-1/2} \hat{\xi_{0}} H^{-1} \rightsquigarrow 0$, the expression for the DM energy density as a function of $a$, the scale parameter, is given by 
\begin{equation}
\Omega_m (a) = \frac{c a}{1 + ca}, \quad \text{with} \quad \eta = (24 \pi G_{N})^{-1/2} \hat{\xi_{0}} H^{-1} \rightsquigarrow 0,
\label{analytic_Y_s_0_eta_0_diss_II}
\end{equation}
being $c$ an integration constant. Accordingly, in the limit $a\rightarrow{0}$, the DM energy density vanishes, which remarks that this era is dominated by radiation with $\omega_{eff}=1/3$. In the other asymptotic region, when $\eta \propto H^{-1} \rightsquigarrow 1$, and $\Omega_m(a) \rightsquigarrow 1$, we obtain
\begin{equation}
\Omega_m (a) = \frac{A a^{4}-3}{A a^{4}+1}, \quad \text{with} \quad \eta = (24 \pi G_{N})^{-1/2} \hat{\xi_{0}} H^{-1} \rightsquigarrow 1,
\label{analytic_Y_s_0_eta_1_diss_II}
\end{equation}
where $A$ is a positive integration constant. The above expression holds in the asymptotic region of large $a$-values, such that $ \Omega_m \sim 1 $. The effective DM viscous pressure can be written as
\begin{equation}
\Pi  = - 3 H \hat{\xi} = - \frac{3^{1/2} \hat{ \xi_0}} {(8\pi G_{N})^{3/2} } ~ H, 
\label{analytic_Pressure_s_0_diss_II}
\end{equation}
which leads to a late accelerated expansion of De Sitter type, as $\eta = (24 \pi G_{N})^{-1/2} \hat{\xi_{0}} H^{-1} \rightsquigarrow 1$, compare with ref. \cite{Padmanabhan:1987dg}.
\\

Similarly to the former analysis, for the particular $s=1$ case, it is also possible to integrate analytically the dynamical system in the asymptotic regime of constant $\eta$. The behavior for the DM energy density parameter as a function of the scale parameter $a$ turns out to be
\begin{equation}
\Omega_m (a) = \frac{a^{1 + 3 \eta}}{1 + a^{1 + 3 \eta}}, \quad \text{with} \quad \eta = (24 \pi G_{N})^{1/2} \hat{\xi_{0}} H \rightsquigarrow 1,
\label{analytic_Y_s_1_diss_II}
\end{equation}
which vanishes in the very early universe as $a\rightarrow{0}$. Moreover, since $\eta \rightsquigarrow 1$, this solution describes a De Sitter inflationary scenario, in agreements with a similar results found in \cite{Padmanabhan:1987dg}. Analogously, for the DM-dominated era $a$ grows exponentially, and therefore $\Omega_m \rightarrow{1} $. This asymptotic behavior is captured by the fixed point of type {I}. As a consequence, the negative effective DM pressure, Eqn.~(\ref{sec2:eqn2}), can be expressed in the form
\begin{equation}
\Pi  = - 3 H \hat{\xi} = - \frac{3^{3/2} H^3 \hat{ \xi_0}} {(8\pi G_{N})^{1/2} } \frac{a^{1 + 3 \eta}}{1 + a^{1 + 3 \eta}}. 
\label{analytic_Pressure_s_1_diss_II}
\end{equation}
As $H$ and $\eta$ are both positive quantities, this effective pressure drives the universe to an accelerated expansion as the scale parameter $a$ grows, while it becomes negligible as $a$ becomes relatively small, compatible with an expected radiation dominated epoch, including the Big Bang singularity ($a \rightarrow{0}$). 
Concerning the limit $\eta \rightarrow 0$, which corresponds to a fixed point of type $II$, the expression for $\Omega_m$ goes continuously to the expression $\Omega_m (a) = a / (1+a)$, for $a$ sufficiently small. Moreover, the asymptotic expression for the effective DM pressure, Eqn.~(\ref{sec2:eqn2}), would become negligible as
\begin{equation}
\Pi (a) \propto {\eta}^3 Y(a) \xrightarrow{\eta \rightarrow{0}} 0, 
\label{analytic_Pressure_s_1_diss_II}
\end{equation} 
which means that close to this type of fixed point, corresponding to a radiation dominated era, the effective pressure would not lead to changes in the dynamics of the early cosmic expansion. Nevertheless, as it was explained in the general discussion of stability properties of fixed points, for $s>1/2$, the stability behavior would correspond to an attractor in the $\eta$-direction, which fails to describe the true physical dynamics of the universe, and hence, this viscous value must be discarded as a suited viscous model. (See the appendix for a detailed explanation on how a radiation dominated fixed point only exists for $0<s<1/2$).  

\section{Conclusions}\label{conclusions}

For the general viscous model $I$, which depends explicitly on the Hubble parameter and the energy density in the form $\xi^I \propto H^{1-2s} \rho_{m}^{s}$, we have found a family of rational values $s = (2n+1)/2m$ with $n, m \in \mathbb{N}$, including the known case $s=1/2$, that are well-suited to describe a complete cosmological evolution. This conclusion is supported by the fixed points associated with both radiation-dominated and viscous dark matter-dominated cosmological eras. Moreover, these suited values have the remarkable feature of displaying an additional fixed point of type $III$, which describes a combined radiation and viscous dark matter behavior and does not lead to an accelerated expansion. We have also observed that in the asymptotic upper limit $s \rightarrow 1_+  $ ($ n=m $ large), this fixed point goes into a fixed point of type $II$, describing standard radiation with an EoS parameter $\omega_{eff} = 1/3$.\\
\\Concerning the unified viscous model $II$, which depends on the dark matter energy density as a power law: $\xi\sim \rho_{m}^{s}$, our dynamical system analysis has revealed significant inconsistencies for negative exponents $s < 0$, which have not been previously observed. Indeed, they must be discarded as suited effective scenarios to describe viscosity within the DM sector, as they fail to produce a radiation-dominated era, an undesirable outcome for any cosmological model. Moreover, in a neighborhood of a rational negative viscous exponent like $ -1/2 $, there will not exist a fixed point representing the radiation dominated era, rendering all these values invalid. This result underscores the need to include this constraint as a prior in statistical analyses of observational data, with implications for current statistical inferences, where both prior and best-fit values of $s$ often fall outside the acceptable range \cite{Bulk2, Yang:2019qza,Yang:2023qqz}.\\
\\ Further, for this type of viscous model, we have found that arbitrary real exponents $s$ lying in the interval $ 0 \leq s < 1/2$ are suited values to model viscosity within the DM sector. They can describe the complete cosmological dynamics, including the known radiation- and dark matter-dominated eras, and lead to an accelerated expansion. Moreover, the stability properties of the associated fixed points also correctly describe the observed universe evolution.\\
\\ The main conclusion of this paper underscores a crucial aspect often overlooked in observational analyses. While early studies hinted at challenges in reconciling model $II$ with observations due to rapid damping of density perturbations in the viscous fluid \cite{Bulk2}, recent analyses incorporating more comprehensive data, including the full cosmic microwave background temperature and anisotropy data, provide strong evidence supporting a non-zero bulk viscous scenario at more than 2 standard deviations \cite{Yang:2019qza,Yang:2023qqz}. These analyses suggest a preferred value of the exponent $s=-0.557$ based on combinations of cosmological datasets. However, the theoretical inconsistencies of this model when $s<0$ raise significant concerns as we have shown. This invalidates the analysis of \cite{Bulk2, Yang:2019qza,Yang:2023qqz} because, initially, this physical information was not included in the prior assumptions.\\
\\ We have observed that in models $\xi^I \propto H^{1-2s} \rho_{m}^{s}$, a natural limit to the standard (pressureless) cold dark matter component is achievable as $s$ approaches infinity asymptotically. However, this feature cannot be recovered for the model $\xi^{II}\sim \rho_{m}^{s}$ unless we trivially choose a vanishing bulk viscosity coefficient $\xi\to 0$.\\
\\ We have shown that for unified models, regardless of the chosen parameterization to model viscosity, the accelerated phase of cosmic expansion requires very high values of bulk viscosity. Such high values are not expected given the very low value of cosmic dark matter density at the present time, and considering the small bulk viscosity values for fluids in terrestrial environments. We pointed out that despite the ongoing debate about the nature of dark matter, it is highly improbable for any model of this component to result in such high values of viscosity. Furthermore, viscosity during structure formation must be very low to fit the data, demanding the counter-intuitive behavior of bulk viscosity diminishing as the energy density increases. In summary, the unified viscous scenario has a weakness of requiring non-physical behavior for the bulk viscosity in terms of fluid density to fit cosmological data. Moreover, the dynamical system analysis of the two models studied reveals that this inverse dependence on matter energy density does not lead to a complete and consistent description of the background cosmic evolution.\\
\\ After exploring predictions and implications for cosmic dynamics, and uncovering some inconsistencies not previously observed, we are left with a fundamental question: Are viscous unified scenarios a viable model to describe the universe? Not surprisingly, the ultimate answer lies in the realm of observational constraints. This is the next step we plan to take in our immediate research.\\

\textbf{Data Availability Statement}: No Data associated in the manuscript.

\section*{Acknowledgments}
G.P. and N. C. acknowledge financial support by Dicyt-USACH Grant No. 042231PA. 
\appendix*
\label{Apendice}

\section{Complementary analysis of stability for the model $II$}
We show explicitly how the behavior described for the model $II$ depends on whether $s \in (0,1/2)$, or $s \in (1/2,1)$. It can be done by choosing different phase space variables given by 
\begin{equation}
x = (8 \pi G_N)^{1/2} H , \quad \text{and} \quad y = \frac{(8 \pi G_N)^2} { 3}  \rho_m. 
\label{new_Ph_space_var_viscous_II_II_b}
\end{equation}
The corresponding evolution equations are:
\begin{align}
\nonumber
&x^\prime= -2 x^2 + \frac{1}{2} y + \frac{3^{s+1/2}}{2} ~ \hat{ \xi_0} ~ x ~ y^{s}, \label{ph_space_par_02a} \\ 
&y^\prime= -\frac{3}{2}~ x ~y + 3^{s+1/2} ~ \hat{ \xi_0} ~ x^2 ~ y^{s}.
\end{align} 
where the $\prime$ stands for the derivative w.r.t. the dimensionless time $\tau = t/(8 \pi G_N)^{1/2}$. The finite fixed points of the above dynamical system are of two types. 

Type $I$ ($s \neq 1/2$):   
\begin{equation}
x_I = \frac{1}{\sqrt{3}} ~\hat{ \xi_0}^\frac {1}{1-2s} , \quad y_I = \frac{1}{3} ~\hat{ \xi_0}^\frac {2}{1-2s}, \quad \text{with} \quad w_{eff}= -1 \quad \text{and} \quad q= -1.  
\label{GFP_02_I} 
\end{equation}

Type $II$ ($s > 0, s \neq 1/2$):
\begin{equation}
x_{II} = 0, \quad y_{II} = 0, \quad \text{with} \quad w_{eff}= 0 \quad \text{and} \quad q= 1/2,
\label{GFP_02_II}
\end{equation}
where we have used the expressions for the EoS parameter $\omega_{eff}$, and for the deceleration parameter $q$ as functions of the phase space variables $(x,y)$, which are given by:

\begin{equation}
w_{eff} = -\frac {2}{3} \frac{x^\prime} {x^2} - 1, \quad \text{and} \quad q = \frac {1}{2} ~ (1 + 3 ~ w_{eff} ),
\label{w_eff_accel_02}
\end{equation}
respectively. It follows that the fixed point of type $I$ corresponds to a viscous DM component, with an EoS parameter associated with negative pressure, leading to accelerated expansion in the late-time dynamics. It is worth pointing out that the dynamical system is singular for $s=1/2$, as there are no associated fixed points for this value.
The eigenvalues associated with the fixed point of type $I$ can straightforwardly be computed from the characteristic equation, leading to the explicit analytic expression
\begin{equation}
\lambda_{\pm} = \left[-\frac{1}{2\sqrt{3}}(11/2 - 3s) \pm \left( \left[\frac{1}{2\sqrt{3}}(11/2 - 3s)\right]^2 + 4(s-1/2) \right)^{1/2}\right]  \left[~\hat{ \xi_0}\right]^\frac {1}{1-2s}. 
\label{viscous_Model_II_par_02_type_I}
\end{equation}
From this expression, one concludes that $s<1/2$ must be required to ensure that this fixed point remains an attractor of the dynamical system. 
As to a fixed point associated to pure radiation ($\rho_m = 0$), even if we restrict $s>0$, it follows from the Eqn. (\ref{ph_space_par_02a}) that $x=0$. Therefore, we conclude that there are no other fixed points in the finite plane $(x,y)$. One can look for fixed points at infinity by using the Poincare sphere projection method, which could describe pure radiation in the same infinite limit (i.e. $x \rightarrow \infty $). Let us, accordingly, introduce the map to the half Poincare sphere using the relations

\begin{equation}
X = \frac {x}{\sqrt{1+x^2+y^2}}, \quad Y = \frac {y}{\sqrt{1+x^2+y^2}} \quad  \text{and} \quad Z = \frac {1}{\sqrt{1+x^2+y^2}}.
\label{Poincare_sphere}
\end{equation}

As it is well known, the topological properties of the original planar flow described by the system of Eqns. (\ref{ph_space_par_02a}) closed to infinity correspond to those of the flow on the Poincare sphere closed to the half-sphere rim at $Z=0$ (see Ref. \cite{Poincare}). For $0<s<1/2$ and close to $Z \sim 0 $, the associated dynamical system turns out to be 

\begin{align}
\nonumber
&Y^\prime= Y Z^{s} + \frac{1}{2}  Y^2 Z^{s+1} + \frac{3^{s+ 1/2}}{2} \hat{ \xi_0} ~ Y^{s} ~(Z-2), \label{Poincare_dyn_system} \\ 
&Z^\prime=  -2 Z^{s+1} + \frac{1}{2}  Y Z^{s+2} + \frac{3^{s+ 1/2}}{2} \hat{ \xi_0} ~ Z^2 ~ Y^{s}.
\end{align} 
From the fixed points of the above system, we are interested in $X=1$, $Y=0$ and $Y =0$, which corresponds to the fixed point $ ( H=\infty,  
 \rho_m = 0)$ of the original system (compare with the analysis of this point below Eqn. (\ref{eigenvalues_s_n_viscous_II_II_b})). This fixed point describes a pure radiation universe era. Unfortunately, it doesn't fulfills  Malkin's theorem around $(Y = 0, Z=0)$ as $s<1/2$. Even worst, the first derivatives at this point are ill-defined. So, the use of this new phase space variables definition, although advantageous to establish the dynamical character of the fixed point $I$, fails to set the stability of the fixed point associated with radiation.  We have then to come back to the original phase space variables (see Eqn.~(\ref{sec3:eqn1})). In this sense, both representations are complementary and fully characterize the dynamical nature of both fixed points.

\bibliography{biblio.bib}

\begin{thebibliography}{54}%
\makeatletter
\providecommand \@ifxundefined [1]{%
 \@ifx{#1\undefined}
}%
\providecommand \@ifnum [1]{%
 \ifnum #1\expandafter \@firstoftwo
 \else \expandafter \@secondoftwo
 \fi
}%
\providecommand \@ifx [1]{%
 \ifx #1\expandafter \@firstoftwo
 \else \expandafter \@secondoftwo
 \fi
}%
\providecommand \natexlab [1]{#1}%
\providecommand \enquote  [1]{``#1''}%
\providecommand \bibnamefont  [1]{#1}%
\providecommand \bibfnamefont [1]{#1}%
\providecommand \citenamefont [1]{#1}%
\providecommand \href@noop [0]{\@secondoftwo}%
\providecommand \href [0]{\begingroup \@sanitize@url \@href}%
\providecommand \@href[1]{\@@startlink{#1}\@@href}%
\providecommand \@@href[1]{\endgroup#1\@@endlink}%
\providecommand \@sanitize@url [0]{\catcode `\\12\catcode `\$12\catcode `\&12\catcode `\#12\catcode `\^12\catcode `\_12\catcode `\%12\relax}%
\providecommand \@@startlink[1]{}%
\providecommand \@@endlink[0]{}%
\providecommand \url  [0]{\begingroup\@sanitize@url \@url }%
\providecommand \@url [1]{\endgroup\@href {#1}{\urlprefix }}%
\providecommand \urlprefix  [0]{URL }%
\providecommand \Eprint [0]{\href }%
\providecommand \doibase [0]{http://dx.doi.org/}%
\providecommand \selectlanguage [0]{\@gobble}%
\providecommand \bibinfo  [0]{\@secondoftwo}%
\providecommand \bibfield  [0]{\@secondoftwo}%
\providecommand \translation [1]{[#1]}%
\providecommand \BibitemOpen [0]{}%
\providecommand \bibitemStop [0]{}%
\providecommand \bibitemNoStop [0]{.\EOS\space}%
\providecommand \EOS [0]{\spacefactor3000\relax}%
\providecommand \BibitemShut  [1]{\csname bibitem#1\endcsname}%
\let\auto@bib@innerbib\@empty
\bibitem [{\citenamefont {Peebles}(1994)}]{Peebles:1994xt}%
  \BibitemOpen
  \bibfield  {author} {\bibinfo {author} {\bibfnamefont {P.~J.~E.}\ \bibnamefont {Peebles}},\ }\href@noop {} {\emph {\bibinfo {title} {{Principles of physical cosmology}}}}\ (\bibinfo {year} {1994})\BibitemShut {NoStop}%
\bibitem [{\citenamefont {Peebles}\ and\ \citenamefont {Ratra}(2003)}]{Peebles:2002gy}%
  \BibitemOpen
  \bibfield  {author} {\bibinfo {author} {\bibfnamefont {P.~J.~E.}\ \bibnamefont {Peebles}}\ and\ \bibinfo {author} {\bibfnamefont {B.}~\bibnamefont {Ratra}},\ }\href {\doibase 10.1103/RevModPhys.75.559} {\bibfield  {journal} {\bibinfo  {journal} {Rev. Mod. Phys.}\ }\textbf {\bibinfo {volume} {75}},\ \bibinfo {pages} {559} (\bibinfo {year} {2003})},\ \Eprint {http://arxiv.org/abs/astro-ph/0207347} {arXiv:astro-ph/0207347} \BibitemShut {NoStop}%
\bibitem [{\citenamefont {Padmanabhan}\ and\ \citenamefont {Chitre}(1987{\natexlab{a}})}]{Padmanabhan:1987dg}%
  \BibitemOpen
  \bibfield  {author} {\bibinfo {author} {\bibfnamefont {T.}~\bibnamefont {Padmanabhan}}\ and\ \bibinfo {author} {\bibfnamefont {S.~M.}\ \bibnamefont {Chitre}},\ }\href {\doibase 10.1016/0375-9601(87)90104-6} {\bibfield  {journal} {\bibinfo  {journal} {Phys. Lett. A}\ }\textbf {\bibinfo {volume} {120}},\ \bibinfo {pages} {433} (\bibinfo {year} {1987}{\natexlab{a}})}\BibitemShut {NoStop}%
\bibitem [{\citenamefont {Fabris}\ \emph {et~al.}(2006)\citenamefont {Fabris}, \citenamefont {Goncalves},\ and\ \citenamefont {de~Sa~Ribeiro}}]{Bulk1}%
  \BibitemOpen
  \bibfield  {author} {\bibinfo {author} {\bibfnamefont {J.~C.}\ \bibnamefont {Fabris}}, \bibinfo {author} {\bibfnamefont {S.~V.~B.}\ \bibnamefont {Goncalves}}, \ and\ \bibinfo {author} {\bibfnamefont {R.}~\bibnamefont {de~Sa~Ribeiro}},\ }\href {\doibase 10.1007/s10714-006-0236-y} {\bibfield  {journal} {\bibinfo  {journal} {Gen. Rel. Grav.}\ }\textbf {\bibinfo {volume} {38}},\ \bibinfo {pages} {495} (\bibinfo {year} {2006})},\ \Eprint {http://arxiv.org/abs/astro-ph/0503362} {arXiv:astro-ph/0503362} \BibitemShut {NoStop}%
\bibitem [{\citenamefont {Li}\ and\ \citenamefont {Barrow}(2009)}]{Bulk2}%
  \BibitemOpen
  \bibfield  {author} {\bibinfo {author} {\bibfnamefont {B.}~\bibnamefont {Li}}\ and\ \bibinfo {author} {\bibfnamefont {J.~D.}\ \bibnamefont {Barrow}},\ }\href {\doibase 10.1103/PhysRevD.79.103521} {\bibfield  {journal} {\bibinfo  {journal} {Phys. Rev. D}\ }\textbf {\bibinfo {volume} {79}},\ \bibinfo {pages} {103521} (\bibinfo {year} {2009})},\ \Eprint {http://arxiv.org/abs/0902.3163} {arXiv:0902.3163 [gr-qc]} \BibitemShut {NoStop}%
\bibitem [{\citenamefont {Colistete}\ \emph {et~al.}(2007)\citenamefont {Colistete}, \citenamefont {Fabris}, \citenamefont {Tossa},\ and\ \citenamefont {Zimdahl}}]{Colistete:2007xi}%
  \BibitemOpen
  \bibfield  {author} {\bibinfo {author} {\bibfnamefont {R.}~\bibnamefont {Colistete}}, \bibinfo {author} {\bibfnamefont {J.~C.}\ \bibnamefont {Fabris}}, \bibinfo {author} {\bibfnamefont {J.}~\bibnamefont {Tossa}}, \ and\ \bibinfo {author} {\bibfnamefont {W.}~\bibnamefont {Zimdahl}},\ }\href {\doibase 10.1103/PhysRevD.76.103516} {\bibfield  {journal} {\bibinfo  {journal} {Phys. Rev. D}\ }\textbf {\bibinfo {volume} {76}},\ \bibinfo {pages} {103516} (\bibinfo {year} {2007})},\ \Eprint {http://arxiv.org/abs/0706.4086} {arXiv:0706.4086 [astro-ph]} \BibitemShut {NoStop}%
\bibitem [{\citenamefont {Brevik}\ \emph {et~al.}(2017)\citenamefont {Brevik}, \citenamefont {Gr\o{}n}, \citenamefont {de~Haro}, \citenamefont {Odintsov},\ and\ \citenamefont {Saridakis}}]{Brevik:2017msy}%
  \BibitemOpen
  \bibfield  {author} {\bibinfo {author} {\bibfnamefont {I.}~\bibnamefont {Brevik}}, \bibinfo {author} {\bibfnamefont {O.}~\bibnamefont {Gr\o{}n}}, \bibinfo {author} {\bibfnamefont {J.}~\bibnamefont {de~Haro}}, \bibinfo {author} {\bibfnamefont {S.~D.}\ \bibnamefont {Odintsov}}, \ and\ \bibinfo {author} {\bibfnamefont {E.~N.}\ \bibnamefont {Saridakis}},\ }\href {\doibase 10.1142/S0218271817300245} {\bibfield  {journal} {\bibinfo  {journal} {Int. J. Mod. Phys. D}\ }\textbf {\bibinfo {volume} {26}},\ \bibinfo {pages} {1730024} (\bibinfo {year} {2017})},\ \Eprint {http://arxiv.org/abs/1706.02543} {arXiv:1706.02543 [gr-qc]} \BibitemShut {NoStop}%
\bibitem [{\citenamefont {Blas}\ \emph {et~al.}(2015)\citenamefont {Blas}, \citenamefont {Floerchinger}, \citenamefont {Garny}, \citenamefont {Tetradis},\ and\ \citenamefont {Wiedemann}}]{blas2015large}%
  \BibitemOpen
  \bibfield  {author} {\bibinfo {author} {\bibfnamefont {D.}~\bibnamefont {Blas}}, \bibinfo {author} {\bibfnamefont {S.}~\bibnamefont {Floerchinger}}, \bibinfo {author} {\bibfnamefont {M.}~\bibnamefont {Garny}}, \bibinfo {author} {\bibfnamefont {N.}~\bibnamefont {Tetradis}}, \ and\ \bibinfo {author} {\bibfnamefont {U.~A.}\ \bibnamefont {Wiedemann}},\ }\href {\doibase 10.1088/1475-7516/2015/11/049} {\bibfield  {journal} {\bibinfo  {journal} {JCAP}\ }\textbf {\bibinfo {volume} {11}},\ \bibinfo {pages} {049} (\bibinfo {year} {2015})},\ \Eprint {http://arxiv.org/abs/1507.06665} {arXiv:1507.06665 [astro-ph.CO]} \BibitemShut {NoStop}%
\bibitem [{\citenamefont {Barbosa}\ \emph {et~al.}(2017)\citenamefont {Barbosa}, \citenamefont {Velten}, \citenamefont {Fabris},\ and\ \citenamefont {Ramos}}]{Barbosa:2017ojt}%
  \BibitemOpen
  \bibfield  {author} {\bibinfo {author} {\bibfnamefont {C.~M.~S.}\ \bibnamefont {Barbosa}}, \bibinfo {author} {\bibfnamefont {H.}~\bibnamefont {Velten}}, \bibinfo {author} {\bibfnamefont {J.~C.}\ \bibnamefont {Fabris}}, \ and\ \bibinfo {author} {\bibfnamefont {R.~O.}\ \bibnamefont {Ramos}},\ }\href {\doibase 10.1103/PhysRevD.96.023527} {\bibfield  {journal} {\bibinfo  {journal} {Phys. Rev. D}\ }\textbf {\bibinfo {volume} {96}},\ \bibinfo {pages} {023527} (\bibinfo {year} {2017})},\ \Eprint {http://arxiv.org/abs/1702.07040} {arXiv:1702.07040 [astro-ph.CO]} \BibitemShut {NoStop}%
\bibitem [{\citenamefont {Foot}\ and\ \citenamefont {Vagnozzi}(2016)}]{foot2016solving}%
  \BibitemOpen
  \bibfield  {author} {\bibinfo {author} {\bibfnamefont {R.}~\bibnamefont {Foot}}\ and\ \bibinfo {author} {\bibfnamefont {S.}~\bibnamefont {Vagnozzi}},\ }\href {\doibase 10.1088/1475-7516/2016/07/013} {\bibfield  {journal} {\bibinfo  {journal} {JCAP}\ }\textbf {\bibinfo {volume} {07}},\ \bibinfo {pages} {013} (\bibinfo {year} {2016})},\ \Eprint {http://arxiv.org/abs/1602.02467} {arXiv:1602.02467 [astro-ph.CO]} \BibitemShut {NoStop}%
\bibitem [{\citenamefont {Gagnon}\ and\ \citenamefont {Lesgourgues}(2011)}]{Gagnon:2011id}%
  \BibitemOpen
  \bibfield  {author} {\bibinfo {author} {\bibfnamefont {J.-S.}\ \bibnamefont {Gagnon}}\ and\ \bibinfo {author} {\bibfnamefont {J.}~\bibnamefont {Lesgourgues}},\ }\href {\doibase 10.1088/1475-7516/2011/09/026} {\bibfield  {journal} {\bibinfo  {journal} {JCAP}\ }\textbf {\bibinfo {volume} {09}},\ \bibinfo {pages} {026} (\bibinfo {year} {2011})},\ \Eprint {http://arxiv.org/abs/1107.1503} {arXiv:1107.1503 [astro-ph.CO]} \BibitemShut {NoStop}%
\bibitem [{\citenamefont {Brevik}\ and\ \citenamefont {Stokkan}(1996)}]{Brevik:1996ca}%
  \BibitemOpen
  \bibfield  {author} {\bibinfo {author} {\bibfnamefont {I.~H.}\ \bibnamefont {Brevik}}\ and\ \bibinfo {author} {\bibfnamefont {G.}~\bibnamefont {Stokkan}},\ }\href {\doibase 10.1007/BF00653769} {\bibfield  {journal} {\bibinfo  {journal} {Astrophys. Space Sci.}\ }\textbf {\bibinfo {volume} {239}},\ \bibinfo {pages} {89} (\bibinfo {year} {1996})}\BibitemShut {NoStop}%
\bibitem [{\citenamefont {Murphy}(1973{\natexlab{a}})}]{Murphy:1973zz}%
  \BibitemOpen
  \bibfield  {author} {\bibinfo {author} {\bibfnamefont {G.~L.}\ \bibnamefont {Murphy}},\ }\href {\doibase 10.1103/PhysRevD.8.4231} {\bibfield  {journal} {\bibinfo  {journal} {Phys. Rev. D}\ }\textbf {\bibinfo {volume} {8}},\ \bibinfo {pages} {4231} (\bibinfo {year} {1973}{\natexlab{a}})}\BibitemShut {NoStop}%
\bibitem [{\citenamefont {Hu}(1982)}]{Hu:1982uea}%
  \BibitemOpen
  \bibfield  {author} {\bibinfo {author} {\bibfnamefont {B.~L.}\ \bibnamefont {Hu}},\ }\href@noop {} {\bibfield  {journal} {\bibinfo  {journal} {Phys. Lett. A}\ }\textbf {\bibinfo {volume} {90}} (\bibinfo {year} {1982})}\BibitemShut {NoStop}%
\bibitem [{\citenamefont {Eshaghi}\ \emph {et~al.}(2015)\citenamefont {Eshaghi}, \citenamefont {Riazi},\ and\ \citenamefont {Kiasatpour}}]{Eshaghi:2015tqa}%
  \BibitemOpen
  \bibfield  {author} {\bibinfo {author} {\bibfnamefont {M.}~\bibnamefont {Eshaghi}}, \bibinfo {author} {\bibfnamefont {N.}~\bibnamefont {Riazi}}, \ and\ \bibinfo {author} {\bibfnamefont {A.}~\bibnamefont {Kiasatpour}},\ }\href@noop {} {\  (\bibinfo {year} {2015})},\ \Eprint {http://arxiv.org/abs/1504.07774} {arXiv:1504.07774 [gr-qc]} \BibitemShut {NoStop}%
\bibitem [{\citenamefont {Bamba}\ and\ \citenamefont {Odintsov}(2016)}]{Bamba:2015sxa}%
  \BibitemOpen
  \bibfield  {author} {\bibinfo {author} {\bibfnamefont {K.}~\bibnamefont {Bamba}}\ and\ \bibinfo {author} {\bibfnamefont {S.~D.}\ \bibnamefont {Odintsov}},\ }\href {\doibase 10.1140/epjc/s10052-015-3861-3} {\bibfield  {journal} {\bibinfo  {journal} {Eur. Phys. J. C}\ }\textbf {\bibinfo {volume} {76}},\ \bibinfo {pages} {18} (\bibinfo {year} {2016})},\ \Eprint {http://arxiv.org/abs/1508.05451} {arXiv:1508.05451 [gr-qc]} \BibitemShut {NoStop}%
\bibitem [{\citenamefont {Wilson}\ \emph {et~al.}(2007)\citenamefont {Wilson}, \citenamefont {Mathews},\ and\ \citenamefont {Fuller}}]{Wilson:2006gf}%
  \BibitemOpen
  \bibfield  {author} {\bibinfo {author} {\bibfnamefont {J.~R.}\ \bibnamefont {Wilson}}, \bibinfo {author} {\bibfnamefont {G.~J.}\ \bibnamefont {Mathews}}, \ and\ \bibinfo {author} {\bibfnamefont {G.~M.}\ \bibnamefont {Fuller}},\ }\href {\doibase 10.1103/PhysRevD.75.043521} {\bibfield  {journal} {\bibinfo  {journal} {Phys. Rev. D}\ }\textbf {\bibinfo {volume} {75}},\ \bibinfo {pages} {043521} (\bibinfo {year} {2007})},\ \Eprint {http://arxiv.org/abs/astro-ph/0609687} {arXiv:astro-ph/0609687} \BibitemShut {NoStop}%
\bibitem [{\citenamefont {Mathews}\ \emph {et~al.}(2008)\citenamefont {Mathews}, \citenamefont {Lan},\ and\ \citenamefont {Kolda}}]{Mathews:2008hk}%
  \BibitemOpen
  \bibfield  {author} {\bibinfo {author} {\bibfnamefont {G.~J.}\ \bibnamefont {Mathews}}, \bibinfo {author} {\bibfnamefont {N.~Q.}\ \bibnamefont {Lan}}, \ and\ \bibinfo {author} {\bibfnamefont {C.}~\bibnamefont {Kolda}},\ }\href {\doibase 10.1103/PhysRevD.78.043525} {\bibfield  {journal} {\bibinfo  {journal} {Phys. Rev. D}\ }\textbf {\bibinfo {volume} {78}},\ \bibinfo {pages} {043525} (\bibinfo {year} {2008})},\ \Eprint {http://arxiv.org/abs/0801.0853} {arXiv:0801.0853 [astro-ph]} \BibitemShut {NoStop}%
\bibitem [{\citenamefont {Atreya}\ \emph {et~al.}(2018)\citenamefont {Atreya}, \citenamefont {Bhatt},\ and\ \citenamefont {Mishra}}]{Atreya:2017pny}%
  \BibitemOpen
  \bibfield  {author} {\bibinfo {author} {\bibfnamefont {A.}~\bibnamefont {Atreya}}, \bibinfo {author} {\bibfnamefont {J.~R.}\ \bibnamefont {Bhatt}}, \ and\ \bibinfo {author} {\bibfnamefont {A.}~\bibnamefont {Mishra}},\ }\href {\doibase 10.1088/1475-7516/2018/02/024} {\bibfield  {journal} {\bibinfo  {journal} {JCAP}\ }\textbf {\bibinfo {volume} {02}},\ \bibinfo {pages} {024} (\bibinfo {year} {2018})},\ \Eprint {http://arxiv.org/abs/1709.02163} {arXiv:1709.02163 [astro-ph.CO]} \BibitemShut {NoStop}%
\bibitem [{\citenamefont {Natwariya}\ \emph {et~al.}(2020)\citenamefont {Natwariya}, \citenamefont {Bhatt},\ and\ \citenamefont {Pandey}}]{Natwariya:2019fif}%
  \BibitemOpen
  \bibfield  {author} {\bibinfo {author} {\bibfnamefont {P.~K.}\ \bibnamefont {Natwariya}}, \bibinfo {author} {\bibfnamefont {J.~R.}\ \bibnamefont {Bhatt}}, \ and\ \bibinfo {author} {\bibfnamefont {A.~K.}\ \bibnamefont {Pandey}},\ }\href {\doibase 10.1140/epjc/s10052-020-8341-8} {\bibfield  {journal} {\bibinfo  {journal} {Eur. Phys. J. C}\ }\textbf {\bibinfo {volume} {80}},\ \bibinfo {pages} {767} (\bibinfo {year} {2020})},\ \Eprint {http://arxiv.org/abs/1907.03445} {arXiv:1907.03445 [astro-ph.CO]} \BibitemShut {NoStop}%
\bibitem [{\citenamefont {Hofmann}\ \emph {et~al.}(2001)\citenamefont {Hofmann}, \citenamefont {Schwarz},\ and\ \citenamefont {Stoecker}}]{Hofmann:2001bi}%
  \BibitemOpen
  \bibfield  {author} {\bibinfo {author} {\bibfnamefont {S.}~\bibnamefont {Hofmann}}, \bibinfo {author} {\bibfnamefont {D.~J.}\ \bibnamefont {Schwarz}}, \ and\ \bibinfo {author} {\bibfnamefont {H.}~\bibnamefont {Stoecker}},\ }\href {\doibase 10.1103/PhysRevD.64.083507} {\bibfield  {journal} {\bibinfo  {journal} {Phys. Rev. D}\ }\textbf {\bibinfo {volume} {64}},\ \bibinfo {pages} {083507} (\bibinfo {year} {2001})},\ \Eprint {http://arxiv.org/abs/astro-ph/0104173} {arXiv:astro-ph/0104173} \BibitemShut {NoStop}%
\bibitem [{\citenamefont {Eckart}(1940)}]{Eckart}%
  \BibitemOpen
  \bibfield  {author} {\bibinfo {author} {\bibfnamefont {C.}~\bibnamefont {Eckart}},\ }\href {\doibase 10.1103/PhysRev.58.919} {\bibfield  {journal} {\bibinfo  {journal} {Phys. Rev.}\ }\textbf {\bibinfo {volume} {58}},\ \bibinfo {pages} {919} (\bibinfo {year} {1940})}\BibitemShut {NoStop}%
\bibitem [{\citenamefont {Landau}\ and\ \citenamefont {Lifshitz}()}]{LandauandLifshitz}%
  \BibitemOpen
  \bibfield  {author} {\bibinfo {author} {\bibfnamefont {L.~D.}\ \bibnamefont {Landau}}\ and\ \bibinfo {author} {\bibfnamefont {E.~M.}\ \bibnamefont {Lifshitz}},\ }\href@noop {} {\bibinfo  {journal} {Fluid Mechanics Volume 6, (1959) Pages 523-529}\ }\BibitemShut {NoStop}%
\bibitem [{\citenamefont {G\'omez}\ \emph {et~al.}(2023)\citenamefont {G\'omez}, \citenamefont {Palma}, \citenamefont {Gonz\'alez}, \citenamefont {Rinc\'on},\ and\ \citenamefont {Cruz}}]{Gomez:2022qcu}%
  \BibitemOpen
\bibfield  {journal} {  }\bibfield  {author} {\bibinfo {author} {\bibfnamefont {G.}~\bibnamefont {G\'omez}}, \bibinfo {author} {\bibfnamefont {G.}~\bibnamefont {Palma}}, \bibinfo {author} {\bibfnamefont {E.}~\bibnamefont {Gonz\'alez}}, \bibinfo {author} {\bibfnamefont {A.}~\bibnamefont {Rinc\'on}}, \ and\ \bibinfo {author} {\bibfnamefont {N.}~\bibnamefont {Cruz}},\ }\href {\doibase 10.1140/epjp/s13360-023-04367-6} {\bibfield  {journal} {\bibinfo  {journal} {Eur. Phys. J. Plus}\ }\textbf {\bibinfo {volume} {138}},\ \bibinfo {pages} {738} (\bibinfo {year} {2023})},\ \Eprint {http://arxiv.org/abs/2210.09429} {arXiv:2210.09429 [gr-qc]} \BibitemShut {NoStop}%
\bibitem [{\citenamefont {Murphy}(1973{\natexlab{b}})}]{Big.Bang}%
  \BibitemOpen
  \bibfield  {author} {\bibinfo {author} {\bibfnamefont {G.~L.}\ \bibnamefont {Murphy}},\ }\href {\doibase 10.1103/PhysRevD.8.4231} {\bibfield  {journal} {\bibinfo  {journal} {Phys. Rev. D}\ }\textbf {\bibinfo {volume} {8}},\ \bibinfo {pages} {4231} (\bibinfo {year} {1973}{\natexlab{b}})}\BibitemShut {NoStop}%
\bibitem [{\citenamefont {Padmanabhan}\ and\ \citenamefont {Chitre}(1987{\natexlab{b}})}]{rho1}%
  \BibitemOpen
  \bibfield  {author} {\bibinfo {author} {\bibfnamefont {T.}~\bibnamefont {Padmanabhan}}\ and\ \bibinfo {author} {\bibfnamefont {S.~M.}\ \bibnamefont {Chitre}},\ }\href {\doibase 10.1016/0375-9601(87)90104-6} {\bibfield  {journal} {\bibinfo  {journal} {Phys. Lett. A}\ }\textbf {\bibinfo {volume} {120}},\ \bibinfo {pages} {433} (\bibinfo {year} {1987}{\natexlab{b}})}\BibitemShut {NoStop}%
\bibitem [{\citenamefont {Brevik}\ and\ \citenamefont {Gorbunova}(2005)}]{Brevik}%
  \BibitemOpen
  \bibfield  {author} {\bibinfo {author} {\bibfnamefont {I.~H.}\ \bibnamefont {Brevik}}\ and\ \bibinfo {author} {\bibfnamefont {O.}~\bibnamefont {Gorbunova}},\ }\href {\doibase 10.1007/s10714-005-0178-9} {\bibfield  {journal} {\bibinfo  {journal} {Gen. Rel. Grav.}\ }\textbf {\bibinfo {volume} {37}},\ \bibinfo {pages} {2039} (\bibinfo {year} {2005})},\ \Eprint {http://arxiv.org/abs/gr-qc/0504001} {arXiv:gr-qc/0504001} \BibitemShut {NoStop}%
\bibitem [{\citenamefont {Cruz}\ \emph {et~al.}(2018)\citenamefont {Cruz}, \citenamefont {Gonz\'alez}, \citenamefont {Lepe},\ and\ \citenamefont {S\'aez-Chill\'on~G\'omez}}]{Analysing}%
  \BibitemOpen
  \bibfield  {author} {\bibinfo {author} {\bibfnamefont {N.}~\bibnamefont {Cruz}}, \bibinfo {author} {\bibfnamefont {E.}~\bibnamefont {Gonz\'alez}}, \bibinfo {author} {\bibfnamefont {S.}~\bibnamefont {Lepe}}, \ and\ \bibinfo {author} {\bibfnamefont {D.}~\bibnamefont {S\'aez-Chill\'on~G\'omez}},\ }\href {\doibase 10.1088/1475-7516/2018/12/017} {\bibfield  {journal} {\bibinfo  {journal} {JCAP}\ }\textbf {\bibinfo {volume} {12}},\ \bibinfo {pages} {017} (\bibinfo {year} {2018})},\ \Eprint {http://arxiv.org/abs/1807.10729} {arXiv:1807.10729 [gr-qc]} \BibitemShut {NoStop}%
\bibitem [{\citenamefont {Normann}\ and\ \citenamefont {Brevik}(2017)}]{valorxi1}%
  \BibitemOpen
  \bibfield  {author} {\bibinfo {author} {\bibfnamefont {B.~D.}\ \bibnamefont {Normann}}\ and\ \bibinfo {author} {\bibfnamefont {I.}~\bibnamefont {Brevik}},\ }\href {\doibase 10.1142/S0217732317500262} {\bibfield  {journal} {\bibinfo  {journal} {Mod. Phys. Lett. A}\ }\textbf {\bibinfo {volume} {32}},\ \bibinfo {pages} {1750026} (\bibinfo {year} {2017})},\ \Eprint {http://arxiv.org/abs/1612.01794} {arXiv:1612.01794 [gr-qc]} \BibitemShut {NoStop}%
\bibitem [{\citenamefont {Normann}\ and\ \citenamefont {Brevik}(2016)}]{valorxi2}%
  \BibitemOpen
  \bibfield  {author} {\bibinfo {author} {\bibfnamefont {B.~D.}\ \bibnamefont {Normann}}\ and\ \bibinfo {author} {\bibfnamefont {I.}~\bibnamefont {Brevik}},\ }\href {\doibase 10.3390/e18060215} {\bibfield  {journal} {\bibinfo  {journal} {Entropy}\ }\textbf {\bibinfo {volume} {18}},\ \bibinfo {pages} {215} (\bibinfo {year} {2016})},\ \Eprint {http://arxiv.org/abs/1601.04519} {arXiv:1601.04519 [gr-qc]} \BibitemShut {NoStop}%
\bibitem [{\citenamefont {Cruz}\ \emph {et~al.}(2022{\natexlab{a}})\citenamefont {Cruz}, \citenamefont {Gonz\'alez},\ and\ \citenamefont {Jovel}}]{primerarticulo}%
  \BibitemOpen
  \bibfield  {author} {\bibinfo {author} {\bibfnamefont {N.}~\bibnamefont {Cruz}}, \bibinfo {author} {\bibfnamefont {E.}~\bibnamefont {Gonz\'alez}}, \ and\ \bibinfo {author} {\bibfnamefont {J.}~\bibnamefont {Jovel}},\ }\href {\doibase 10.1103/PhysRevD.105.024047} {\bibfield  {journal} {\bibinfo  {journal} {Phys. Rev. D}\ }\textbf {\bibinfo {volume} {105}},\ \bibinfo {pages} {024047} (\bibinfo {year} {2022}{\natexlab{a}})},\ \Eprint {http://arxiv.org/abs/2109.09865} {arXiv:2109.09865 [gr-qc]} \BibitemShut {NoStop}%
\bibitem [{\citenamefont {Cruz}\ \emph {et~al.}(2022{\natexlab{b}})\citenamefont {Cruz}, \citenamefont {Gonz\'alez},\ and\ \citenamefont {Jovel}}]{sym14091866}%
  \BibitemOpen
  \bibfield  {author} {\bibinfo {author} {\bibfnamefont {N.}~\bibnamefont {Cruz}}, \bibinfo {author} {\bibfnamefont {E.}~\bibnamefont {Gonz\'alez}}, \ and\ \bibinfo {author} {\bibfnamefont {J.}~\bibnamefont {Jovel}},\ }\href {\doibase 10.3390/sym14091866} {\bibfield  {journal} {\bibinfo  {journal} {Symmetry}\ }\textbf {\bibinfo {volume} {14}} (\bibinfo {year} {2022}{\natexlab{b}}),\ 10.3390/sym14091866}\BibitemShut {NoStop}%
\bibitem [{\citenamefont {Cruz}\ \emph {et~al.}(2022{\natexlab{c}})\citenamefont {Cruz}, \citenamefont {Gonz\'alez},\ and\ \citenamefont {Jovel}}]{Cruz:2022zxe}%
  \BibitemOpen
  \bibfield  {author} {\bibinfo {author} {\bibfnamefont {N.}~\bibnamefont {Cruz}}, \bibinfo {author} {\bibfnamefont {E.}~\bibnamefont {Gonz\'alez}}, \ and\ \bibinfo {author} {\bibfnamefont {J.}~\bibnamefont {Jovel}},\ }\href@noop {} {\  (\bibinfo {year} {2022}{\natexlab{c}})},\ \Eprint {http://arxiv.org/abs/2207.14244} {arXiv:2207.14244 [gr-qc]} \BibitemShut {NoStop}%
\bibitem [{\citenamefont {Coley}\ and\ \citenamefont {van~den Hoogen}(1995)}]{Coley:1995dpj}%
  \BibitemOpen
  \bibfield  {author} {\bibinfo {author} {\bibfnamefont {A.~A.}\ \bibnamefont {Coley}}\ and\ \bibinfo {author} {\bibfnamefont {R.~J.}\ \bibnamefont {van~den Hoogen}},\ }\href {\doibase 10.1088/0264-9381/12/8/015} {\bibfield  {journal} {\bibinfo  {journal} {Class. Quant. Grav.}\ }\textbf {\bibinfo {volume} {12}},\ \bibinfo {pages} {1977} (\bibinfo {year} {1995})},\ \Eprint {http://arxiv.org/abs/gr-qc/9605061} {arXiv:gr-qc/9605061} \BibitemShut {NoStop}%
\bibitem [{\citenamefont {Coley}\ \emph {et~al.}(1996)\citenamefont {Coley}, \citenamefont {van~den Hoogen},\ and\ \citenamefont {Maartens}}]{coley1996qualitative}%
  \BibitemOpen
  \bibfield  {author} {\bibinfo {author} {\bibfnamefont {A.~A.}\ \bibnamefont {Coley}}, \bibinfo {author} {\bibfnamefont {R.~J.}\ \bibnamefont {van~den Hoogen}}, \ and\ \bibinfo {author} {\bibfnamefont {R.}~\bibnamefont {Maartens}},\ }\href {\doibase 10.1103/PhysRevD.54.1393} {\bibfield  {journal} {\bibinfo  {journal} {Phys. Rev. D}\ }\textbf {\bibinfo {volume} {54}},\ \bibinfo {pages} {1393} (\bibinfo {year} {1996})},\ \Eprint {http://arxiv.org/abs/gr-qc/9605063} {arXiv:gr-qc/9605063} \BibitemShut {NoStop}%
\bibitem [{\citenamefont {Avelino}\ and\ \citenamefont {Nucamendi}(2009)}]{Bulk4}%
  \BibitemOpen
  \bibfield  {author} {\bibinfo {author} {\bibfnamefont {A.}~\bibnamefont {Avelino}}\ and\ \bibinfo {author} {\bibfnamefont {U.}~\bibnamefont {Nucamendi}},\ }\href {\doibase 10.1088/1475-7516/2009/04/006} {\bibfield  {journal} {\bibinfo  {journal} {JCAP}\ }\textbf {\bibinfo {volume} {04}},\ \bibinfo {pages} {006} (\bibinfo {year} {2009})},\ \Eprint {http://arxiv.org/abs/0811.3253} {arXiv:0811.3253 [gr-qc]} \BibitemShut {NoStop}%
\bibitem [{\citenamefont {{Acquaviva}}\ and\ \citenamefont {{Beesham}}(2015)}]{acquaviva2015nonlinear}%
  \BibitemOpen
  \bibfield  {author} {\bibinfo {author} {\bibfnamefont {G.}~\bibnamefont {{Acquaviva}}}\ and\ \bibinfo {author} {\bibfnamefont {A.}~\bibnamefont {{Beesham}}},\ }\href {\doibase 10.1088/0264-9381/32/21/215026} {\bibfield  {journal} {\bibinfo  {journal} {Class. Quantum Grav.}\ }\textbf {\bibinfo {volume} {32}},\ \bibinfo {eid} {215026} (\bibinfo {year} {2015})}\BibitemShut {NoStop}%
\bibitem [{\citenamefont {Acquaviva}\ and\ \citenamefont {Beesham}(2014)}]{Acquaviva:2014vga}%
  \BibitemOpen
  \bibfield  {author} {\bibinfo {author} {\bibfnamefont {G.}~\bibnamefont {Acquaviva}}\ and\ \bibinfo {author} {\bibfnamefont {A.}~\bibnamefont {Beesham}},\ }\href {\doibase 10.1103/PhysRevD.90.023503} {\bibfield  {journal} {\bibinfo  {journal} {Phys. Rev. D}\ }\textbf {\bibinfo {volume} {90}},\ \bibinfo {pages} {023503} (\bibinfo {year} {2014})},\ \Eprint {http://arxiv.org/abs/1405.3459} {arXiv:1405.3459 [gr-qc]} \BibitemShut {NoStop}%
\bibitem [{\citenamefont {Avelino}\ \emph {et~al.}(2013{\natexlab{a}})\citenamefont {Avelino}, \citenamefont {Leyva},\ and\ \citenamefont {Urena-Lopez}}]{Avelino:2013wea}%
  \BibitemOpen
  \bibfield  {author} {\bibinfo {author} {\bibfnamefont {A.}~\bibnamefont {Avelino}}, \bibinfo {author} {\bibfnamefont {Y.}~\bibnamefont {Leyva}}, \ and\ \bibinfo {author} {\bibfnamefont {L.~A.}\ \bibnamefont {Urena-Lopez}},\ }\href {\doibase 10.1103/PhysRevD.88.123004} {\bibfield  {journal} {\bibinfo  {journal} {Phys. Rev. D}\ }\textbf {\bibinfo {volume} {88}},\ \bibinfo {pages} {123004} (\bibinfo {year} {2013}{\natexlab{a}})},\ \Eprint {http://arxiv.org/abs/1306.3270} {arXiv:1306.3270 [astro-ph.CO]} \BibitemShut {NoStop}%
\bibitem [{\citenamefont {Avelino}\ \emph {et~al.}(2013{\natexlab{b}})\citenamefont {Avelino}, \citenamefont {Garcia-Salcedo}, \citenamefont {Gonzalez}, \citenamefont {Nucamendi},\ and\ \citenamefont {Quiros}}]{Avelino:2013mua}%
  \BibitemOpen
  \bibfield  {author} {\bibinfo {author} {\bibfnamefont {A.}~\bibnamefont {Avelino}}, \bibinfo {author} {\bibfnamefont {R.}~\bibnamefont {Garcia-Salcedo}}, \bibinfo {author} {\bibfnamefont {T.}~\bibnamefont {Gonzalez}}, \bibinfo {author} {\bibfnamefont {U.}~\bibnamefont {Nucamendi}}, \ and\ \bibinfo {author} {\bibfnamefont {I.}~\bibnamefont {Quiros}},\ }\href {\doibase 10.1088/1475-7516/2013/08/012} {\bibfield  {journal} {\bibinfo  {journal} {JCAP}\ }\textbf {\bibinfo {volume} {08}},\ \bibinfo {pages} {012} (\bibinfo {year} {2013}{\natexlab{b}})},\ \Eprint {http://arxiv.org/abs/1303.5167} {arXiv:1303.5167 [gr-qc]} \BibitemShut {NoStop}%
\bibitem [{\citenamefont {Sasidharan}\ and\ \citenamefont {Mathew}(2016)}]{Sasidharan:2015ihq}%
  \BibitemOpen
  \bibfield  {author} {\bibinfo {author} {\bibfnamefont {A.}~\bibnamefont {Sasidharan}}\ and\ \bibinfo {author} {\bibfnamefont {T.~K.}\ \bibnamefont {Mathew}},\ }\href {\doibase 10.1007/JHEP06(2016)138} {\bibfield  {journal} {\bibinfo  {journal} {JHEP}\ }\textbf {\bibinfo {volume} {06}},\ \bibinfo {pages} {138} (\bibinfo {year} {2016})},\ \Eprint {http://arxiv.org/abs/1511.05287} {arXiv:1511.05287 [gr-qc]} \BibitemShut {NoStop}%
\bibitem [{\citenamefont {Cruz}\ \emph {et~al.}(2023)\citenamefont {Cruz}, \citenamefont {Gomez}, \citenamefont {Gonzalez}, \citenamefont {Palma},\ and\ \citenamefont {Rincon}}]{Cruz:2023dzn}%
  \BibitemOpen
  \bibfield  {author} {\bibinfo {author} {\bibfnamefont {N.}~\bibnamefont {Cruz}}, \bibinfo {author} {\bibfnamefont {G.}~\bibnamefont {Gomez}}, \bibinfo {author} {\bibfnamefont {E.}~\bibnamefont {Gonzalez}}, \bibinfo {author} {\bibfnamefont {G.}~\bibnamefont {Palma}}, \ and\ \bibinfo {author} {\bibfnamefont {A.}~\bibnamefont {Rincon}},\ }\href {\doibase 10.1016/j.dark.2023.101351} {\bibfield  {journal} {\bibinfo  {journal} {Phys. Dark Univ.}\ }\textbf {\bibinfo {volume} {42}},\ \bibinfo {pages} {101351} (\bibinfo {year} {2023})},\ \Eprint {http://arxiv.org/abs/2304.12407} {arXiv:2304.12407 [gr-qc]} \BibitemShut {NoStop}%
\bibitem [{\citenamefont {Aghanim}\ \emph {et~al.}(2020)\citenamefont {Aghanim} \emph {et~al.}}]{Planck:2018vyg}%
  \BibitemOpen
  \bibfield  {author} {\bibinfo {author} {\bibfnamefont {N.}~\bibnamefont {Aghanim}} \emph {et~al.} (\bibinfo {collaboration} {Planck}),\ }\href {\doibase 10.1051/0004-6361/201833910} {\bibfield  {journal} {\bibinfo  {journal} {Astron. Astrophys.}\ }\textbf {\bibinfo {volume} {641}},\ \bibinfo {pages} {A6} (\bibinfo {year} {2020})},\ \bibinfo {note} {[Erratum: Astron.Astrophys. 652, C4 (2021)]},\ \Eprint {http://arxiv.org/abs/1807.06209} {arXiv:1807.06209 [astro-ph.CO]} \BibitemShut {NoStop}%
\bibitem [{\citenamefont {Pavón}\ and\ \citenamefont {Zimdahl}(1993)}]{PAVON1993261}%
  \BibitemOpen
  \bibfield  {author} {\bibinfo {author} {\bibfnamefont {D.}~\bibnamefont {Pavón}}\ and\ \bibinfo {author} {\bibfnamefont {W.}~\bibnamefont {Zimdahl}},\ }\href {\doibase https://doi.org/10.1016/0375-9601(93)90675-P} {\bibfield  {journal} {\bibinfo  {journal} {Physics Letters A}\ }\textbf {\bibinfo {volume} {179}},\ \bibinfo {pages} {261} (\bibinfo {year} {1993})}\BibitemShut {NoStop}%
\bibitem [{\citenamefont {Most}\ \emph {et~al.}(2022)\citenamefont {Most}, \citenamefont {Haber}, \citenamefont {Harris}, \citenamefont {Zhang}, \citenamefont {Alford},\ and\ \citenamefont {Noronha}}]{Most:2022yhe}%
  \BibitemOpen
  \bibfield  {author} {\bibinfo {author} {\bibfnamefont {E.~R.}\ \bibnamefont {Most}}, \bibinfo {author} {\bibfnamefont {A.}~\bibnamefont {Haber}}, \bibinfo {author} {\bibfnamefont {S.~P.}\ \bibnamefont {Harris}}, \bibinfo {author} {\bibfnamefont {Z.}~\bibnamefont {Zhang}}, \bibinfo {author} {\bibfnamefont {M.~G.}\ \bibnamefont {Alford}}, \ and\ \bibinfo {author} {\bibfnamefont {J.}~\bibnamefont {Noronha}},\ }\href@noop {} {\  (\bibinfo {year} {2022})},\ \Eprint {http://arxiv.org/abs/2207.00442} {arXiv:2207.00442 [astro-ph.HE]} \BibitemShut {NoStop}%
\bibitem [{\citenamefont {Sharma}\ \emph {et~al.}(2023)\citenamefont {Sharma}, \citenamefont {Pareek},\ and\ \citenamefont {Kumar}}]{sharma2023bulk}%
  \BibitemOpen
  \bibfield  {author} {\bibinfo {author} {\bibfnamefont {B.}~\bibnamefont {Sharma}}, \bibinfo {author} {\bibfnamefont {S.}~\bibnamefont {Pareek}}, \ and\ \bibinfo {author} {\bibfnamefont {R.}~\bibnamefont {Kumar}},\ }\href@noop {} {\bibfield  {journal} {\bibinfo  {journal} {European Journal of Mechanics-B/Fluids}\ }\textbf {\bibinfo {volume} {98}},\ \bibinfo {pages} {32} (\bibinfo {year} {2023})}\BibitemShut {NoStop}%
\bibitem [{\citenamefont {Di~Valentino}\ \emph {et~al.}(2021)\citenamefont {Di~Valentino}, \citenamefont {Mena}, \citenamefont {Pan}, \citenamefont {Visinelli}, \citenamefont {Yang}, \citenamefont {Melchiorri}, \citenamefont {Mota}, \citenamefont {Riess},\ and\ \citenamefont {Silk}}]{DiValentino:2021izs}%
  \BibitemOpen
  \bibfield  {author} {\bibinfo {author} {\bibfnamefont {E.}~\bibnamefont {Di~Valentino}}, \bibinfo {author} {\bibfnamefont {O.}~\bibnamefont {Mena}}, \bibinfo {author} {\bibfnamefont {S.}~\bibnamefont {Pan}}, \bibinfo {author} {\bibfnamefont {L.}~\bibnamefont {Visinelli}}, \bibinfo {author} {\bibfnamefont {W.}~\bibnamefont {Yang}}, \bibinfo {author} {\bibfnamefont {A.}~\bibnamefont {Melchiorri}}, \bibinfo {author} {\bibfnamefont {D.~F.}\ \bibnamefont {Mota}}, \bibinfo {author} {\bibfnamefont {A.~G.}\ \bibnamefont {Riess}}, \ and\ \bibinfo {author} {\bibfnamefont {J.}~\bibnamefont {Silk}},\ }\href {\doibase 10.1088/1361-6382/ac086d} {\bibfield  {journal} {\bibinfo  {journal} {Class. Quant. Grav.}\ }\textbf {\bibinfo {volume} {38}},\ \bibinfo {pages} {153001} (\bibinfo {year} {2021})},\ \Eprint {http://arxiv.org/abs/2103.01183} {arXiv:2103.01183 [astro-ph.CO]} \BibitemShut {NoStop}%
\bibitem [{\citenamefont {Bento}\ \emph {et~al.}(2002)\citenamefont {Bento}, \citenamefont {Bertolami},\ and\ \citenamefont {Sen}}]{Bento:2002ps}%
  \BibitemOpen
  \bibfield  {author} {\bibinfo {author} {\bibfnamefont {M.~C.}\ \bibnamefont {Bento}}, \bibinfo {author} {\bibfnamefont {O.}~\bibnamefont {Bertolami}}, \ and\ \bibinfo {author} {\bibfnamefont {A.~A.}\ \bibnamefont {Sen}},\ }\href {\doibase 10.1103/PhysRevD.66.043507} {\bibfield  {journal} {\bibinfo  {journal} {Phys. Rev. D}\ }\textbf {\bibinfo {volume} {66}},\ \bibinfo {pages} {043507} (\bibinfo {year} {2002})},\ \Eprint {http://arxiv.org/abs/gr-qc/0202064} {arXiv:gr-qc/0202064} \BibitemShut {NoStop}%
\bibitem [{\citenamefont {Yang}\ \emph {et~al.}(2019)\citenamefont {Yang}, \citenamefont {Pan}, \citenamefont {Di~Valentino}, \citenamefont {Paliathanasis},\ and\ \citenamefont {Lu}}]{Yang:2019qza}%
  \BibitemOpen
  \bibfield  {author} {\bibinfo {author} {\bibfnamefont {W.}~\bibnamefont {Yang}}, \bibinfo {author} {\bibfnamefont {S.}~\bibnamefont {Pan}}, \bibinfo {author} {\bibfnamefont {E.}~\bibnamefont {Di~Valentino}}, \bibinfo {author} {\bibfnamefont {A.}~\bibnamefont {Paliathanasis}}, \ and\ \bibinfo {author} {\bibfnamefont {J.}~\bibnamefont {Lu}},\ }\href {\doibase 10.1103/PhysRevD.100.103518} {\bibfield  {journal} {\bibinfo  {journal} {Phys. Rev. D}\ }\textbf {\bibinfo {volume} {100}},\ \bibinfo {pages} {103518} (\bibinfo {year} {2019})},\ \Eprint {http://arxiv.org/abs/1906.04162} {arXiv:1906.04162 [astro-ph.CO]} \BibitemShut {NoStop}%
\bibitem [{\citenamefont {{Malkin I. G.}}()}]{Malkin_52}%
  \BibitemOpen
  \bibfield  {author} {\bibinfo {author} {\bibnamefont {{Malkin I. G.}}},\ }\href@noop {} {\bibinfo  {journal} {Gos. Izdat. tekh.-teoret. Lit., Moskow, (1952). English transl., AEC tr-3352 (1958)}\ }\BibitemShut {NoStop}%
\bibitem [{\citenamefont {Bemfica}\ \emph {et~al.}(2023)\citenamefont {Bemfica}, \citenamefont {Disconzi}, \citenamefont {Noronha},\ and\ \citenamefont {Scherrer}}]{Bemfica:2022dnk}%
  \BibitemOpen
\bibfield  {journal} {  }\bibfield  {author} {\bibinfo {author} {\bibfnamefont {F.~S.}\ \bibnamefont {Bemfica}}, \bibinfo {author} {\bibfnamefont {M.~M.}\ \bibnamefont {Disconzi}}, \bibinfo {author} {\bibfnamefont {J.}~\bibnamefont {Noronha}}, \ and\ \bibinfo {author} {\bibfnamefont {R.~J.}\ \bibnamefont {Scherrer}},\ }\href {\doibase 10.1103/PhysRevD.107.023512} {\bibfield  {journal} {\bibinfo  {journal} {Phys. Rev. D}\ }\textbf {\bibinfo {volume} {107}},\ \bibinfo {pages} {023512} (\bibinfo {year} {2023})},\ \Eprint {http://arxiv.org/abs/2210.13372} {arXiv:2210.13372 [gr-qc]} \BibitemShut {NoStop}%
\bibitem [{\citenamefont {Yang}\ \emph {et~al.}(2023)\citenamefont {Yang}, \citenamefont {Pan}, \citenamefont {Di~Valentino}, \citenamefont {Escamilla-Rivera},\ and\ \citenamefont {Paliathanasis}}]{Yang:2023qqz}%
  \BibitemOpen
  \bibfield  {author} {\bibinfo {author} {\bibfnamefont {W.}~\bibnamefont {Yang}}, \bibinfo {author} {\bibfnamefont {S.}~\bibnamefont {Pan}}, \bibinfo {author} {\bibfnamefont {E.}~\bibnamefont {Di~Valentino}}, \bibinfo {author} {\bibfnamefont {C.}~\bibnamefont {Escamilla-Rivera}}, \ and\ \bibinfo {author} {\bibfnamefont {A.}~\bibnamefont {Paliathanasis}},\ }\href {\doibase 10.1093/mnras/stad115} {\bibfield  {journal} {\bibinfo  {journal} {Mon. Not. Roy. Astron. Soc.}\ }\textbf {\bibinfo {volume} {520}},\ \bibinfo {pages} {1146} (\bibinfo {year} {2023})},\ \Eprint {http://arxiv.org/abs/2301.03969} {arXiv:2301.03969 [astro-ph.CO]} \BibitemShut {NoStop}%
\bibitem [{\citenamefont {Acquaviva}\ and\ \citenamefont {Beesham}(2018)}]{Acquaviva:2018rqi}%
  \BibitemOpen
  \bibfield  {author} {\bibinfo {author} {\bibfnamefont {G.}~\bibnamefont {Acquaviva}}\ and\ \bibinfo {author} {\bibfnamefont {A.}~\bibnamefont {Beesham}},\ }\href {\doibase 10.1088/1361-6382/aadb38} {\bibfield  {journal} {\bibinfo  {journal} {Class. Quant. Grav.}\ }\textbf {\bibinfo {volume} {35}},\ \bibinfo {pages} {195011} (\bibinfo {year} {2018})},\ \Eprint {http://arxiv.org/abs/1808.09202} {arXiv:1808.09202 [gr-qc]} \BibitemShut {NoStop}%
\bibitem [{\citenamefont {Perko}(2000)}]{Poincare}%
  \BibitemOpen
  \bibfield  {author} {\bibinfo {author} {\bibfnamefont {L.}~\bibnamefont {Perko}},\ }\href@noop {} {\emph {\bibinfo {title} {{Differential Equations and Dynamical Systems.}}}}\ (\bibinfo  {publisher} {Springer Verlag},\ \bibinfo {address} {New York, Berlin, Heidelberg},\ \bibinfo {year} {(2000)})\BibitemShut {NoStop}%
\end{thebibliography}%

\end{document}